\documentclass[12pt,a4]{article}
\usepackage{latexsym,epsfig,amssymb,euscript,amsmath,verbatim,cite}
\newcommand{\be}{\begin{equation}}
\newcommand{\ee}{\end{equation}}
\newcommand{\bea}{\begin{eqnarray}}
\newcommand{\eea}{\end{eqnarray}}

\newcommand{\refe}[1]{(\ref{#1})}
\newcommand{\nn} {\nonumber}

\newcommand{\ov}{\overline}
\numberwithin{equation}{section}
\usepackage{bm}

\begin{document}
\pagestyle{empty}
\begin{flushright}
{\small{
IPPP/12/79\\
DCPT/12/158}}
\end{flushright}

\begin{center}
{\LARGE{\bf Field Theory Interpretation\\ 
\medskip
of ${\cal N}=2$ Stringy Instantons}}

\vspace{1cm}

{\large{Riccardo Argurio,$^{a,c,}$\footnote{\tt rargurio@ulb.ac.be, $^2$\tt dforcell@ulb.ac.be,
$^3$\tt alberto.mariotti@durham.ac.uk, \\
\phantom{a}\hspace{-.1cm}\phantom{a}
$^4$\tt dmusso@ulb.ac.be, $^5$\tt christoffer.petersson@ulb.ac.be} 
Davide Forcella,$^{a,c,2}$
\\
Alberto Mariotti,$^{b,c,d,3}$
Daniele Musso$^{a,c,e,4}$
\\ and Christoffer Petersson$^{a,c,5}$
\\[1cm]}}

{\small{
{}$^a$ Physique Th\'eorique et Math\'ematique\\
Universit\'e Libre de Bruxelles, C.P. 231, 1050 Bruxelles, Belgium\\
\medskip
{}$^b$ Theoretische Natuurkunde\\
Vrije Universiteit Brussel, Pleinlaan 2, 1050 Brussels, Belgium\\
\medskip
{}$^c$  International Solvay Institutes, Brussels, Belgium\\
\medskip
{}$^d$  Institute for Particle Physics Phenomenology,\\
Department of Physics, Durham University, DH1 3LE, United Kingdom\\
\medskip
{}$^e$  Dipartimento di Fisica, Universit\`a di Genova,\\
via Dodecaneso 33, I-16146, Genova, Italy
}}

\vspace{1cm}

{\bf Abstract}
\end{center}

\noindent
We consider stringy instanton contributions to the prepotential of 
low-energy \mbox{${\cal N}=2$} theories engineered by D-brane set ups at
orbifold and orientifold singularities. We show that such
contributions can always be reproduced in a purely field theoretic UV
completion. We perform the explicit check up to instanton number three,
both for $Sp(0)$ and $U(1)$ stringy instantons, the latter being introduced for
the purpose. We further argue that the UV completion that we propose,
though weakly coupled, can be smoothly mapped to a gravity-dual
inspired UV completion consisting of a cascade of baryonic root transitions.



\newpage

\setcounter{page}{1} \pagestyle{plain} \renewcommand{\thefootnote}{\arabic{footnote}} \setcounter{footnote}{0}

\tableofcontents

\section{Introduction}
A particular class of non-perturbative effects in string theory can be
described by D-brane instantons, i.e.~by the effects on the low-energy
degrees of freedom caused by the presence of the world-volume of an
Euclidean D-brane wrapped on some (compact) cycle of the internal
geometry
\cite{Witten:1995gx,Douglas:1995bn,Ganor:1996pe,Green:1997tv}. When
the low-energy theory comprises gauge theory degrees of freedom,
associated with spacetime-filling (ordinary) D-branes, the effects of
D-brane instantons can sometimes be straightforwardly linked to gauge
theory instantons. String theory actually provides in this case a
somewhat simpler rationale for the ADHM construction \cite{Atiyah:1978ri}, giving a 
physical interpretation for the many fermionic and bosonic zero modes
that appear in that construction (see for instance
\cite{Dorey:2002ik,Bianchi:2007ft}). 

However, there are some instances where D-brane
instantons provide non-perturbative corrections which cannot be
ascribed to ordinary gauge theory instantons
\cite{Blumenhagen:2006xt,Ibanez:2006da,Florea:2006si}. This kind of
non-gauge instanton effects in gauge theories, derived in string
theory by D-brane set ups, have been dubbed stringy
instantons (for a review, see \cite{Blumenhagen:2009qh}). As a rule of
thumb, given a stack of spacetime-filling D-branes, gauge theory
instantons correspond to D-brane instantons lying 
on top of it, while stringy instantons originate from D-brane
instantons orthogonal to it in the compact space.\footnote{
Analogous String theory instanton effects, 
so-called residual instantons, have been 
studied in \cite{Aganagic:2003xq,Intriligator:2003xs}
in the context of matrix models.
For a discussion about the relation between stringy instantons and 
residual instantons see \cite{GarciaEtxebarria:2008iw}.}

Corrections due to stringy instantons were studied mostly focusing on
D-brane set ups leading to gauge theories with ${\cal N}=1$
supersymmetry, the instantons contributing to the superpotential of
the theory. It was soon realized \cite{Argurio:2007vqa,Bianchi:2007wy} that stringy
instantons do actually contribute to the superpotential only in a few
restricted cases, due to the generic presence of some extra neutral
fermionic zero modes. A simple way to get rid of these embarrassing
zero modes, devised in \cite{Argurio:2007vqa} in simple non-compact orbifold
models but easily generalizable, is to project them out by introducing
an orientifold. Another setting is to place a D-brane instanton on top
of an additional single spacetime-filling D-brane \cite{Petersson:2007sc} (see
also \cite{Aganagic:2007py,GarciaEtxebarria:2007zv}), so that the extra zero modes are soaked
up just as in the usual gauge-instanton set up, except for the fact
that for the $U(1)$ theory living on the extra brane there is no
corresponding gauge theory instanton. We remark that in
both of these cases of $\mathcal{N}=1$
stringy instantons, only the single instanton
contributes to the superpotential.

That stringy instantons are really stringy effects was put into
question in \cite{Aharony:2007pr}. There it was shown that, by
embedding the low-energy theory in a cascade of Seiberg dualities (as
is usual in quiver gauge theories arising at Calabi-Yau
singularities), the stringy instanton contribution to the
superpotential could alternatively be derived from known
non-perturbative effects in the (strongly coupled) gauge theory one
step up in the cascade. This kind of field theory interpretation of
stringy instanton superpotential contributions was extended in
\cite{Krefl:2008gs,Amariti:2008xu} to virtually all cases, including both $Sp(0)$
(i.e.~orientifolded) and $U(1)$ stringy instantons. 

Stringy instanton contributions to four-dimensional ${\cal N}=2$
gauge theories have relatively had little attention, possibly due to
less phenomenological interest. They are however of clear
theoretical interest. 
In \cite{Ghorbani:2010ks,Ghorbani:2011xh,Musso:2012sn} they were studied in a
simple orientifold of 
an orbifold set up, thus pertaining to stringy instantons of the
$Sp(0)$ kind (see also \cite{Argurio:2007vqa} for a very brief
glance at the same effect). As it befits an ${\cal N}=2$ set up, they
contribute to the prepotential, and contrarily to the ${\cal N}=1$
case, they contribute for every instanton number. In other words, in
this instance multi-stringy instantons contribute and one can even
hope to resum their contributions. 

Once the existence of ${\cal N}=2$ stringy instantons is established,
it is legitimate to ask whether also for them a field theory
interpretation can be given. Though from the point of view of string
theory, ${\cal N}=2$ stringy instantons are not too different from
their ${\cal N}=1$ cousins, in the field theoretic UV completion
matters could be quite different. Indeed, the AdS/CFT correspondence 
suggests two rather different UV field theory completions. 
In the ${\cal N}=1$ case the completion relied on the persistence of the superpotential through a
Seiberg duality. In the ${\cal N}=2$ case both the superpotential and the
Seiberg duality are not applicable concepts anymore. As it was shown
in \cite{Benini:2008ir} (see also \cite{Cremonesi:2009hq}), 
a low energy D-brane set up emerges rather
from a cascade of baryonic root transitions \cite{Argyres:1996eh}.

In this paper, we show that given a low-energy D-brane set up where
${\cal N}=2$ stringy instantons contribute to the prepotential, there
indeed is a UV completion such that the same contributions can be given
a field theory interpretation, namely in terms of a
prepotential derived from ordinary instanton localization techniques or
from a Seiberg-Witten curve. 
We will check this up to
three-instanton contributions, and both for $Sp(0)$ and $U(1)$ stringy
instantons (which we introduce for the purpose). We also show that the UV
completion, though it does not imply any strong coupling transition,
can be smoothly related to a theory that appears higher up in a
cascade of baryonic root transitions, thus making the proposed UV
completion natural even as a quiver-inspired completion.

The paper is structured as follows. In Section 2 we present the
simplest instance of an ${\cal N}=2$ instanton, and give its field
theory interpretation through a weakly coupled UV completion. Our aim
is to present our main result devoid of all the technicalities
involved in both pictures (stringy and field theoretic). In Section 3
we discuss in much more detail the ${\cal N}=2$ orientifolded case, i.e.~the $Sp(0)$ stringy instantons. Building on
the results of \cite{Ghorbani:2010ks,Ghorbani:2011xh,Musso:2012sn} on the string
side, 
we provide a matching of the two interpretations up to instanton
number three, finding complete agreement. In Section 4 we introduce
the ${\cal N}=2$ version of the $U(1)$ stringy instantons in an
orbifold set up, showing that they indeed contribute to the
prepotential. We then proceed also in this case to exactly match the
results to a specific UV completion. Moreover we discuss some simple 
generalizations of the $U(1)$ orbifolds, and the embedding of our procedure 
into duality cascades. In the Appendices we collect some technical results.
In Appendix \ref{sp_mass} we describe the introduction of masses in the symplectic case. In Appendix \ref{Nekrasov} we provide some 
details about the computations of the gauge instanton contributions using
localization techniques. In Appendix \ref{regu} we discuss three different regularizations 
for the stringy instanton computation and, eventually, in Appendix \ref{analogy} we illustrate the relation between the procedure we use in the main text and the
${\cal N}=1$ set ups. 

\section{A simple ${\cal N}=2$ stringy instanton and its field theory interpretation}
The purpose of this section is to sketch in a very simple example the idea that we will then discuss in all detail and apply to all the occurrences of 
${\cal N}=2$ stringy instantons. 

Our simplest example is provided by fractional D3-branes on an
orientifold of the orbifold $\mathbb{C}^2/\mathbb{Z}_3$, see Figure \ref{sympa} in the next Section
and consider $N_1=N'$ and $N_2=N$. The
four-dimensional ${\cal N}=2$ quiver gauge theory associated to a
general configuration is given by vector multiplets associated to a
gauge group $Sp(N')\times U(N)$, hypermultiplets associated to
bifundamental matter between the two gauge groups and a
hypermultiplet in the symmetric
representation attached to the unitary gauge group.  When there are no
fractional branes along the orientifold plane we have $N'=0$ and the
gauge theory reduces to a single $U(N)$ gauge group with a
symmetric hypermultiplet. 

We now consider the D-brane instantons. As for the spacetime-filling
branes, there are two fractional D$(-1)$-instantons, one on the node
along the orientifold and one on the other node. The first would
correspond to a gauge instanton of the $Sp(N')$ gauge group, with
symmetry group $SO(k)$ for $k$ instantons \cite{Gimon:1996rq}, 
while the second is a gauge
instanton for the $U(N)$ gauge group, with symmetry group $U(k')$.  
When $N'=0$, the instantons on the first node
appear as having a purely stringy origin,
since the $Sp(0)$ gauge theory is empty. However, their $SO(k)$
symmetry group is such that the fermionic zero modes are just the
right number in order to contribute to
the prepotential. 

Let us specialize to a single instanton on top  of the $Sp(0)$
node. One can show
\cite{Argurio:2007vqa,Ghorbani:2010ks,Ghorbani:2011xh} that as a
result of integrating over all the fermionic and bosonic zero modes,
except the four bosonic and the four fermionic zero modes corresponding to the
integration over the chiral ${\cal N}=2$ superspace, the following
contribution to the prepotential is generated:
\be
{\cal F}_\mathrm{1-inst}^{\, s}=M_s^{2-N}e^{2\pi i \tau_D}
\det \Phi ,
\label{f1string}
\ee
where $\Phi$ is the adjoint vector multiplet of the $U(N)$ gauge
group, the superscript $s$ reminds us that we are dealing
with a stringy instanton, $M_s=1/\sqrt{\alpha '}$ is the string mass, and $\tau_D$ is related to the (stringy) volume of the cycle
wrapped by the fractional 
D$(-1)$-instanton. This contribution cannot clearly be attributed
neither to gauge instantons of the $U(N)$ gauge group, nor obviously to ones of
the $Sp(0)$ gauge group. Nevertheless, string theory instructs us that
if we are to consider this gauge theory within this framework, we have
to take into account this instanton correction. Is the latter purely
stringy in nature, then?

We now show that there is another, purely field theoretic, UV
completion of the present gauge theory, that reproduces exactly the
same instanton correction to the prepotential. Consider first the
$U(N)$ gauge group as weakly coupled, essentially providing for a
global ``flavor'' symmetry. Actually, to be precise, in a 
$Sp(N')\times U(N)$ quiver the $Sp(N')$ gauge group has a total of
$N$ fundamental hypermultiplets. Starting then with the $Sp(0)\times U(N)$
theory, we UV complete it with a $Sp(M)\times U(N+M)$ theory,\footnote{We adopt
the notation for which $Sp(1)\simeq SU(2)$. Hence the fundamental
representation of $Sp(M)$ has $2M$ components.} where $M$ of the $Sp(M)$
fundamental hypers have explicit mass terms $m_i$, with $i=1\dots M$. 

Now, we want to turn on vacuum expectation values to the adjoint
scalar of the $Sp(M)$ vector multiplet in such a way that, at low
energies, the effective theory will contain no massless degrees of
freedom except for a charged hypermultiplet for each one of the Cartan
generators of $Sp(M)$, and the spectator symmetric hyper of
$U(N)$. This low-energy theory is reminiscent, and indeed inspired to,
the description of the quantum locus of the root of the baryonic
branch for ${\cal N}=2$ gauge theories with flavors \cite{Argyres:1996eh,Argyres:1996hc}.
In order to obtain this spectrum, we have to choose each of the $M$
masses $m_i$ 
to be different from each other, and then tune the eigenvalues
of the adjoint $\Phi'$ of $Sp(M)$ such that $\phi'_i=m_i$. We can now
compute the one instanton correction to the prepotential in 
the $Sp(M)$ gauge theory in this
particular vacuum, with flavor masses given as above and
complemented by the diagonal values $\phi$ of the scalar adjoint of
$U(N)$, which effectively act as masses for the remaining $N$ light
flavors.

One can use a direct ADHM construction \cite{Atiyah:1978ri} (see also
\cite{Dorey:2002ik}), possibly using the modern approach of
\cite{Nekrasov:2002qd,Marino:2004cn,Nekrasov:2004vw,Shadchin:2004yx,Shadchin:2005mx}. Another approach is to determine
the Seiberg-Witten curve of the theory, from which one extracts the
one-instanton contribution to the prepotential
\cite{D'Hoker:1996nv,D'Hoker:1996mu,
Edelstein:1998sp,Edelstein:1999dd,
Chan:1999gj,Ennes:1999fb}.

In either way, one finds
\be
{\cal
  F}_\mathrm{1-inst}^{\, g}=\frac{\Lambda^{M+2-N}}{\det m}
\det \Phi , 
\label{f1ft}
\ee
where again $\Phi$ is the vector multiplet of the $U(N)$ gauge group,
whose lowest component is
$\phi$, the superscript ``$g$'' reminds us that we are dealing with a
gauge instanton, $\Lambda$ is the dynamical scale of the $Sp(M)$ theory and we
have defined $\det m=\prod_i m_i$. 

We find remarkable agreement between the expressions \refe{f1string}
and \refe{f1ft}, being both a non-trivial prepotential term
proportional to $\det \Phi$. The constant of proportionality can be
made to coincide by just tuning the arbitrary parameters of the field
theory and stringy UV completions, i.e. $m$ and $\Lambda$ against
$M_s$ and $\tau_D$. What we will show in the following, is that once
this tuning is done for the one-instanton contribution, the two and
three gauge instanton contributions exactly coincide with the two and three stringy instanton contributions 
 without the need of any further tuning. 

This is our main result, on which we will elaborate in the rest of
this paper. The stringy instanton contributions to the prepotential
can be exactly reproduced by a simple field theory UV completion,
involving massive flavors and a Higgsed gauge group, and going to a
particular point of the moduli space reminiscent of the baryonic root.

\section{Gauge vs stringy instantons in symplectic gauge theory}

We consider the $\mathcal{N}=2$ $Sp(N') \times U(N)$ model
introduced in the preceding Section.
As anticipated,
the stringy and ordinary gauge
instantons to be compared arise within the non-perturbative sectors 
of two distinct field theories. The ordinary instantons emerge in a gauge theory
which represents an appropriate field theoretic UV completion of the theory affected by the stringy
contributions. We henceforth refer to the former as the \emph{UV theory}
and to the latter as the \emph{IR theory}.

In the following Subsections we first review the computation yielding
the stringy instanton contributions in the IR theory. 
We then construct the UV gauge theory and  
compute the corresponding instanton corrections. Finally we 
successfully compare the instanton contributions in the two descriptions
 up to instanton number three.

\subsection{Stringy instantons for ${\cal N}=2$ $Sp(0)\times U(N)$ theory}

The ${\cal N}=2$ $Sp(0)\times U(N)$ gauge theory appears as the
low-energy field theory description of the open string sector of the
following set up.

\begin{figure}[ht]
  \centering
  \includegraphics[scale=.29]{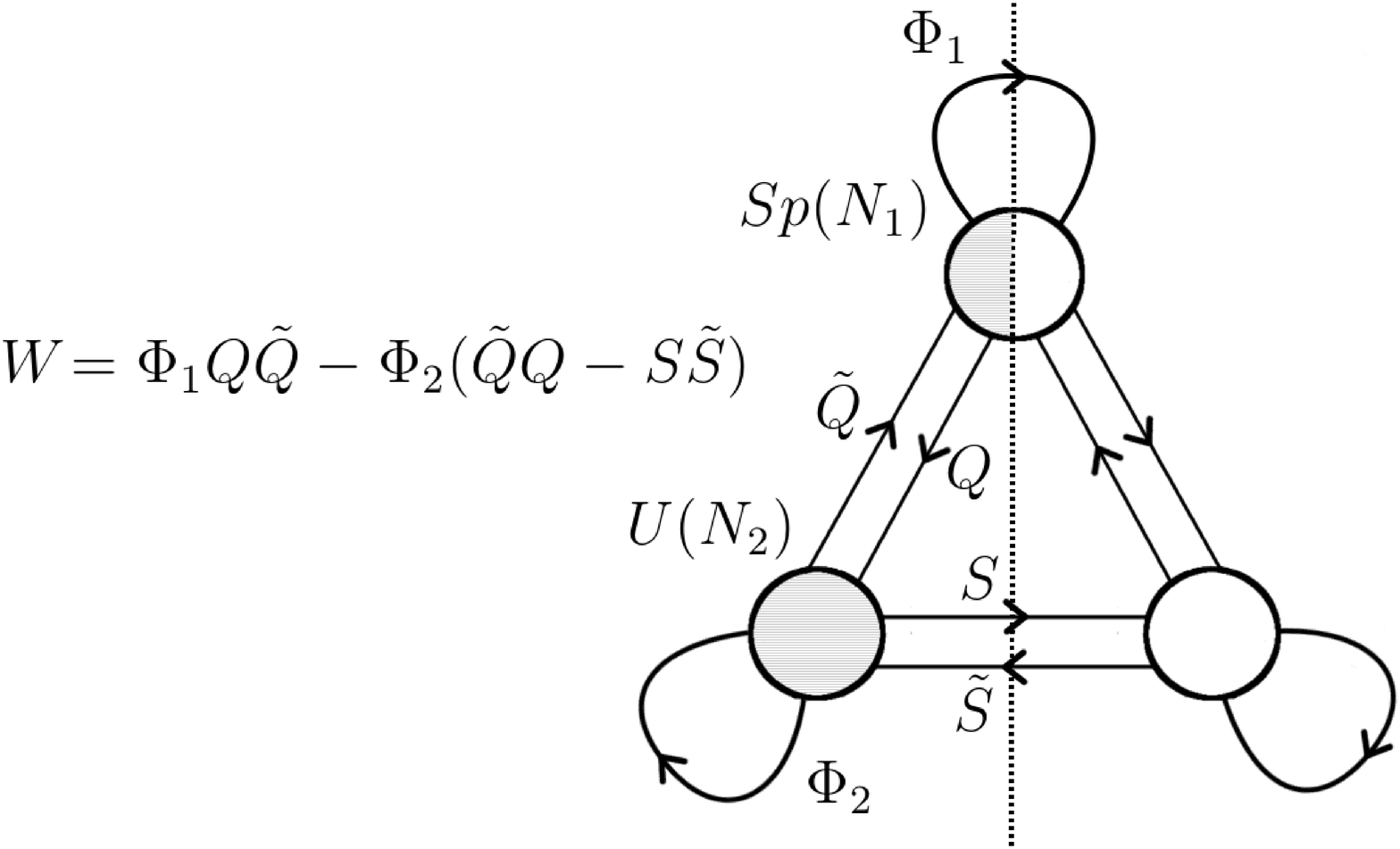}
  \caption{Quiver diagram for the orientifolded $\mathbb{C}^2/\mathbb{Z}_3\times\mathbb{C}$ orbifold.
  The fields $Q$, $\tilde{Q}$  transform in the bifundamental representation of the two gauge nodes while
  the fields $S$, $\tilde{S}$ transform in the symmetric representation of $U(N_2)$.}
  \label{sympa}
\end{figure}

The D-brane system lies on an orientifolded $\mathbb{C}^2/\mathbb{Z}_3\times \mathbb{C}$ 
orbifold background. Before considering the orientifold projection,
the orbifold leads to a quiver with three independent, unitary gauge
factors corresponding to the three irreducible representations of
the $\mathbb{Z}_3$ orbifold group, each defining a fractional D3
brane. After enforcing the orientifold
projection, the model is reduced to a two-node quiver with one unitary
and one symplectic gauge factors. The symplectic factor arises from the
projection of one of the original unitary nodes while the unitary
factor of the projected model emerges from the identification of the 
two other unitary factors of the original quiver (see Figure \ref{sympa}).
The original bi-fundamental matter associated to the string modes stretching
between the two identified unitary nodes is projected down to
matter in the symmetric  representation of the single unitary factor
of the projected quiver. 

The ${\cal N}=2$ $Sp(0)\times U(N)$ gauge theory corresponds to a
configuration with $N$ fractional D3 branes on
the unitary node and none on the symplectic node. However, we consider 
the latter node to be populated by D$(-1)$-instantons. Since the instanton
branes are located on a different node with respect to the D3  
branes, they correspond to stringy instantons.

The $\mathbb{C}^2/\mathbb{Z}_3\times \mathbb{C}$ orientifold background
preserves ${\cal N}=2$ supersymmetry and the direct instantonic
computations are performed within the localization framework for
stringy instantons \cite{Billo:2009di}. From the explicit evaluation
of the instanton partition functions, one is able to compute directly the stringy
instanton contributions to the prepotential of the theory order by
order in the instanton number.
The entire computation was performed in detail in \cite{Ghorbani:2010ks,Ghorbani:2011xh,Musso:2012sn}
of which we simply report the results of use here. 

The stringy,
non-perturbative correction to the prepotential is of the form
\be
\mathcal{F}^{s}= \sum_{k=1}^{\infty} \mathcal{F}^{s}_k =
 \sum_{k=1}^{\infty}
\left( M_s^{2-N} e^{2 \pi i \tau_D} \right)^k F_k^s\ ,
\ee
where the superscript $s$ stands for ``stringy'' and $\mathcal{F}^{s}_k$ is the $k$-th stringy instanton contribution.
The explicit contributions 
at instanton level
$k=1,2$ and $3$ are respectively:%
\footnote{In \cite{Ghorbani:2011xh} the $3$-instanton prepotential lacks 
the factor $(-1)^N$; we repeated and corrected the computation obtaining \eqref{sF3}.}
\begin{eqnarray}\label{sF1}
 F_1^s &=& (-1)^N\ {\cal N}_1\ \text{det}(\Phi_2)\ ,\\ \label{sF2}
 F_2^s &=& \frac{{\cal N}_1^2}{8}\ e_{N-1}(\Phi_2^2)\ ,\\ \label{sF3}
 F_3^s &=& (-1)^N\ \frac{{\cal N}_1^3}{12}\ \text{det}(\Phi_2)\ e_{N-2}(\Phi_2^2) \ ;
\end{eqnarray}
where ${\cal N}_1$ is an overall numerical normalization, 
$\Phi_2$ represents the scalar belonging to the gauge vector multiplet
of the $U(N)$ node
and eventually $e_i(X)$ is the $i$-th symmetric polynomial defined
as follows
\begin{equation}
 e_i(X) = e_i(X_1,...,X_{N}) = \sum_{1\leq j_1 < j_2 < ... < j_i \leq N} X_{j_1} X_{j_2} ... X_{j_i} \ , \label{edef}
\end{equation}
where the $X_i$ are the eigenvalues of $X$.

We note here that the results above can be reproduced using the
approach of
\cite{Nekrasov:2002qd,Marino:2004cn,Nekrasov:2004vw,Shadchin:2004yx,Shadchin:2005mx}
in the case of $Sp(N')$ gauge theories with fundamental matter,
taking however the final expression to be valid even after
continuation to the unphysical case of $Sp(0)$. 

What we are left now to do, is to show that the expressions
\refe{sF1}--\refe{sF3} can be reproduced by an instanton computation
in a physically sensible (i.e. asymptotically free) $Sp(N')$ gauge
theory. The non-trivial task is to match both the field-dependence of
each $k$-instanton contribution, and their normalization which implies
just a single constant for all contributions. 

\subsection{UV and IR symplectic theories}
\label{Spflow}
We now describe the gauge theory UV completion of the $\mathcal{N}=2$
$Sp(0)\times U(N)$  model.
Such UV completion is given by  
the $Sp(N_c)\times U(N_f)$ theory (namely the model corresponding to the  
quiver depicted in Figure \ref{sympa} with
$N_1=N_c$ and $N_2=N_f$)
considered at a specific point in the moduli space.
Indeed, in order to break completely the $Sp(N_c)$ gauge group, we
consider the following diagonal VEV assignment
\begin{equation}\label{VEV}
 {\Phi_1} = (x_1 M,...,x_{N_c}M)\ ,
\end{equation}
where $\Phi_1$ is now the vector whose components parametrize the $N_c$-dimensional
Cartan subgroup of $Sp(N_c)$; $M$ represents a mass scale and the $x$'s are $N_c$
numbers of order one inserted to break the degeneracy among the components of $\Phi_1$.
The VEV's \eqref{VEV} break the original $Sp(N_c)$ gauge group down
to $U(1)^{N_c}$.

We assign masses to $N_c$ out of $N_f$ flavors requiring the existence of $N_c$ massless
hypermultiplets, each of them charged under one of the broken $U(1)^{N_c}$ 
gauge factors (see Appendix \ref{sp_mass}). The mass matrix is then diagonal,
\begin{equation}
 \hat{M} = \big(x_1 M, ..., x_{N_c} M, \underbrace{0,...,0}_{N_f-N_c}\big) \ .
\end{equation}
This choice guarantees the existence of a Higgs branch for each of the
$U(1)^{N_c}$ factors.
The mass assignment preserves a $U(N)$ symmetry with 
\begin{equation}
 N = N_f-N_c\ .
\end{equation}
We can parametrize the small masses of the remaining flavors with $N$ fields $\phi_{2,i}$,
\begin{equation}\label{sp_mass_par}
 \hat{M} = \big(x_1 M, ..., x_{N_c} M, - \phi_{2,1},...,- \phi_{2,N}\big) \ .
\end{equation}
The fields $\phi_{2,i}$ are the eigenvalues of a field $\Phi_2$ in  the adjoint representation of the preserved $U(N)$
flavor group, where we have chosen the negative signs for later convenience.

The IR theory is obtained considering the UV model with the VEV's \eqref{VEV}
and masses \eqref{sp_mass_par} in the large $M$ limit.
The IR theory has therefore an $Sp(0) \times U(N)$ symmetry.

\subsection{Ordinary gauge instantons in $Sp$ gauge theories}

The ordinary gauge instanton contributions to an ${\cal N}=2$ $Sp(N_c)$
gauge theory with matter can be computed following two different approaches:
either by direct calculation through localization techniques 
\cite{Shadchin:2004yx,Shadchin:2005mx}
or by studying the corresponding Seiberg-Witten curve \cite{Ennes:1999fb}.
The two possibilities are equivalent and we adopt the former approach.
In Appendix \ref{Nekrasov}
we briefly sketch the direct calculation described in \cite{Shadchin:2004yx,Shadchin:2005mx}
to which we refer for any further detail. 

The gauge instanton contribution to the prepotential can be expressed as
\be
\label{FTinst}
\mathcal{F}^g= \sum_{k=1}^{\infty} \mathcal{F}^g_k =  \sum_{k=1}^{\infty} F_k^g \Lambda^{k b_0}
\ee
where $\Lambda$ is the strong coupling scale of the $Sp(N_c)$ gauge group,
the superscript $g$ reminds us that we are considering gauge instantons, $\mathcal{F}^g_k$ is the $k$-th gauge instanton contribution,
and the $1$-loop coefficient of the $\beta$-function is given by
\begin{equation}
 b_0=2N_c+2 -N_f\ .
\end{equation}
We report here the first three gauge instanton corrections to the prepotential
\begin{eqnarray}\label{effe1}
 F_1^g &=& -\frac{1}{2} \left(\prod_{l=1}^{N_c} \frac{1}{\phi_{1,l}^2}\right) \left(\prod_{j=1}^{N_f} m_j\right)\ ,\\ \label{effe2}
 F_2^g &=& -\frac{1}{16} \left[ \left( \sum_{l=1}^{N_c} \frac{S_l(\phi_{1,l})}{\phi_{1,l}^4}\right) + \frac{1}{4} \left.\frac{\partial^2 T(t)}{\partial t^2}\right|_{t=0} \right]\ ,\\ \label{effe3}
 F_3^g &=& -\frac{1}{16} \left(\prod_{l=1}^{N_c} \frac{1}{\phi_{1,l}^2}\right) \left(\prod_{j=1}^{N_f} m_j\right)
 \left[\left( \sum_{l=1}^{N_c} \frac{S_{1,l}(\phi_{1,l})}{\phi_{1,l}^6} \right)
 + \frac{1}{144} \left.\frac{\partial^4 T(t)}{\partial t^4}\right|_{t=0} \right]\ ,
\end{eqnarray}
where we have defined 
\begin{equation}\label{reS}
  S_l(t) = \frac{1}{4 \phi_{1,l}\, t} \prod_{k\neq l} \frac{1}{\left(t^2 - \phi_{1,k}^2\right)^2}
  \prod_{j=1}^{N_f} \left(m_j^2 - t^2\right)\ ,
\end{equation}
\begin{equation}\label{T}
 T(t) = \frac{1}{P(t)^2} \prod_{j=1}^{N_f} \left(m_j^2 - t^2\right)\ , \qquad 
  P(t) = \prod_{l=1}^{N_c} (t^2 - \phi_{1,l}^2)\ .
\end{equation}
Note that here $\phi_{1,l}$ denotes the entries of the adjoint field $\Phi_1$ given in (\ref{VEV})
and $m_j$ denotes the entries of the mass array $\hat{M}$ given in \eqref{sp_mass_par}.
The latter includes the diagonal part of the field $\Phi_2$.

\subsection{Comparing gauge and stringy instantons}

In this Subsection we show that the gauge instanton and stringy instanton computations give the same contribution to the prepotential at least up to 
3 instantons: namely ${\cal F}_k^s = {\cal F}_k^g$ for $k=1,2,3$.

In order to compare the gauge instanton contributions in the UV theory
with the stringy instanton contributions in the IR theory, we first
evaluate the gauge instanton corrections to the UV theory with VEV and mass
assignments \eqref{VEV}, \eqref{sp_mass_par} and then study the large $M$ limit.
Notice once again that the mass scale $M$ appears both in the VEV's and masses.

Plugging the VEV's \eqref{VEV} and the masses \eqref{sp_mass_par} into the residue 
function $S$ defined explicitly in \eqref{reS}, we observe that 
\begin{equation}
 S_l(\phi_{1,l}) = 0\ ,
\end{equation}
for any $\phi_{1,l}$ with $l=1,...,N_c$; indeed, the last product in \eqref{reS}
contains always a vanishing factor.
As a consequence, all the terms in $S_l(\phi_{1,l})$ contained in the expression for
the prepotential \eqref{effe2} and \eqref{effe3} vanish too.

\subsubsection{1-instanton contribution}

We start comparing the one stringy instanton contribution with the one gauge instanton contribution.

We consider the expression for the $1$-instanton prepotential given in \eqref{effe1}
plugging into it the VEV's and masses \eqref{VEV}, \eqref{sp_mass_par},
\begin{equation}\label{sp_ord_1}
 F_1^g =- \frac{1}{2} \frac{\ (-1)^{N}\ \phi_{2,1} \cdot ... \cdot \phi_{2,N}}{(x_1 \cdot ... \cdot x_{N_c}) M^{N_c}}
     = - \frac{1}{2}  \frac{\ (-1)^{N}\ \text{det}(\Phi_2)}{\text{det}(x)\ M^{N_c}} \ ,
\end{equation}
where we have introduced the following notation
\begin{eqnarray}
 \text{det}(\Phi_2) &=& \phi_{2,1} \cdot ... \cdot \phi_{2,N}\ , \\
 \text{det}(x) &=& x_1 \cdot ... \cdot x_{N_c} \ .
\end{eqnarray}
Note that the dimensionful prefactor of the gauge instanton prepotential, including
the determinant $\text{det}(x) \,M^{N_c}$,
can be combined in an IR scale 
\be
\label{matching}
\Lambda_{\text{IR}}^{2- N} \equiv \frac{\Lambda_{\text{UV}}^{2 N_c+2 - N_f}}{\text{det}(x) M^{N_c}}\ .
\ee

The $k=1$ gauge instanton correction \eqref{sp_ord_1} 
to the prepotential can be exactly matched with the
stringy instanton one \eqref{sF1}. 
Indeed, we identify the IR theory scale with the 
string parameter $M_s^{2-N} e^{2 \pi i \tau_D}$,
and we fix the overall normalization constant ${\cal N}_1$ 
as follows
\begin{equation}\label{normalization}
\Lambda_{\text{IR}}^{2- N}=M_s^{2-N} e^{2 \pi i \tau_D}
\qquad 
 {\cal N}_1 =- \frac{1}{2} \ .
\end{equation}
In this way, the $k=1$ gauge instanton contribution \eqref{sp_ord_1} equals the
$1$-instanton stringy contribution \eqref{sF1}, namely
\begin{equation}
M_s^{2-N} e^{2 \pi i \tau_D}
F_1^s = \Lambda_{\text{UV}}^{2 N_c+2 - N_f} F_1^g\ ,
\end{equation}
or equivalently
\begin{equation}
{\cal F}_1^s = {\cal F}_1^g\ .
\end{equation}

\subsubsection{2-instanton contribution}
The comparison of the 1-instanton contributions has allowed us to match completely the scales and the normalization factor ${\cal N}_1$ (which could actually be reabsorbed into a redefinition of the stringy scale).  
Now that we have no parameters left to fix, 
let us compare the two instanton computations.

Plugging the VEV and mass assignment \eqref{VEV}, \eqref{sp_mass_par}
into the $2$-instanton prepotential \eqref{effe2}, we simplify $F_2^g$ to
\begin{equation}\label{prep_2}
 F_2^g = -\frac{1}{64} \frac{\partial^2 T(t)}{\partial t^2}\Big|_{t=0} \ ;
\end{equation}
indeed, as already observed, the terms containing the residue function
$S_l(\phi_{1,l})$ vanish. 
We then compute the second derivative of
the $T$ function defined in \eqref{T},
inserting the VEV's \eqref{VEV} and the masses \eqref{sp_mass_par}; for large values of $M$ we obtain
\begin{equation}
 \begin{split}
 F_2^g &= \frac{1}{32} \left(\prod_{l=1}^{N_c} \frac{1}{\phi_{1,l}^4}\right)  \sum_{k=N_c+1}^{N_f} 
 \prod^{N_f}_{j\neq k}
 \ m_j^2 + ...\\
     &= \frac{1}{32} \frac{e_{N-1}(\phi_{2,1}^2,...,\phi_{2,N}^2)}{M^{2N_c}\, \text{det}(x)^2}+ ...
     = \frac{1}{32} \frac{e_{N-1}(\Phi_2^2)}{M^{2N_c}\, \text{det}(x)^2}+ ...
 \end{split}
\end{equation}
where the dots indicate the omission of terms which are subleading 
in the large $M$ limit.

We consider again the matching of the scale (\ref{matching}) and its identification 
with the
string parameter 
\eqref{normalization}
as already done for the $1$-instanton contribution,
including the fixed value for the
overall numerical normalization constant ${\cal N}_1$.
With this mapping,
the $k=2$ gauge instanton 
contribution to the UV theory then
equals the $2$-instanton stringy contribution to the IR model (\ref{sF2}),
\begin{equation}
\left( 
M_s^{2-N} e^{2 \pi i \tau_D}
\right)^2
F_2^s = \Lambda_{\text{UV}}^{2(2 N_c+2 - N_f)} F_2^g\ ,
\end{equation}
or equivalently
\begin{equation}
{\cal F}_2^s = {\cal F}_2^g\ .
\end{equation}

\subsubsection{3-instanton contribution}

Let us now pass to the three instantons case.

We insert the masses \eqref{sp_mass_par} and the VEV's \eqref{VEV} into the $3$-instanton prepotential \eqref{effe3};
in this case as well only the term in $T$ will contribute, so we have
\begin{equation}\label{p3}
 F_3^g = -\frac{1}{16} \left(\prod_{l=1}^N \frac{1}{\phi_{1,l}^2}\right) \left(\prod_{j=1}^{N_f} m_j\right)
 \frac{1}{144} \left.\frac{\partial^4 T(t)}{\partial t^4}\right|_{t=0}\ .
\end{equation}
Using the definition of the function $T(t)$ given in \eqref{T}, 
we compute the relevant terms in its fourth derivative and, in the limit of large $M$, we finally obtain 
\begin{equation}\label{F3S}
 \begin{split}
 F_3^g &= -\frac{1}{16 \cdot 12} \left(\prod_{l=1}^{N_c} \frac{1}{\phi_{1,l}^6}\right) \left(\prod_{h=1}^{N_f} m_h\right)
  \left(  \sum_{k=N_c+1}^{N_f} \sum_{i\neq k}^{N_f} \
 \prod_{j\neq k, \ j\neq i}^{N_f} m_j^2 \right) + ...\\
  &=- \frac{1}{16 \cdot 12}\, \frac{(-1)^{N}\ \text{det}(\Phi_2)}{M^{3 N_c} \text{det}(x)^3 }\
  2\; e_{N-2}(\phi_{2,1}^2,...,\phi_{2,N}^2) + ...\\
  &= -\frac{1}{8 \cdot 12}\, \frac{(-1)^{N}\ \text{det}(\Phi_2)}{M^{3 N_c} \text{det}(x)^3 }\
  e_{N-2}(\Phi_2^2) + ...\ ,
  \end{split}
\end{equation}
where again the dots indicate terms which are suppressed in the large $M$ limit.

Upon considering the scale matching \eqref{matching} 
and the mapping \eqref{normalization}
as in the $k=1$ 
case,
the $k=3$ gauge instanton contribution to the UV theory equals the
stringy $3$-instanton contribution \eqref{sF3} computed directly in the IR theory,
\begin{equation}
\left( 
M_s^{2-N} e^{2 \pi i \tau_D}
\right)^3
F_3^s = \Lambda_{\text{UV}}^{3(2 N_c+2 - N_f)} F_3^g\ ,
\end{equation}
that is
\begin{equation}
{\cal F}_3^s = {\cal F}_3^g\ .
\end{equation}
We hope to have convinced the reader of the matching of the gauge and stringy computation at any instanton number $k$. In Section \ref{cascadeD}, we will give some additional 
arguments to support this conjecture.


\section{Gauge vs stringy instantons in unitary gauge theory}

In this Section we consider a different kind of low-energy theories
that have stringy instanton corrections, namely theories with a $U(1)$
gauge group. In this case, the gauge group is not completely absent as
in the $Sp(0)$ case, however it is Abelian and does not support gauge
instantons. 

We thus consider   
 $\mathcal{N}=2$ SQCD quiver gauge theories realized as worldvolume theories of fractional D$3$-branes probing
an orbifold singularity. 
We first compute the contribution to the prepotential in a $U(1)\times U(N)$ gauge theory arising from D$(-1)$-instantons located at the $U(1)$ node.
We then describe how this configuration can be realized as the IR limit of a UV completed field theory considered
at a specific point on the moduli space.
The ordinary gauge theory instanton corrections in this UV field theory are studied and successfully matched to the 
stringy D$(-1)$-instanton corrections we compute directly in the IR theory. 
We conclude the Section by discussing how the connection between the IR and
UV theory can be naturally understood in terms of a duality cascade.

\subsection{D-instantons in ${\cal N}=2$ Abelian gauge theories}
\label{Uinst}

\begin{figure}[ht]
  \centering
  \includegraphics[scale=.29]{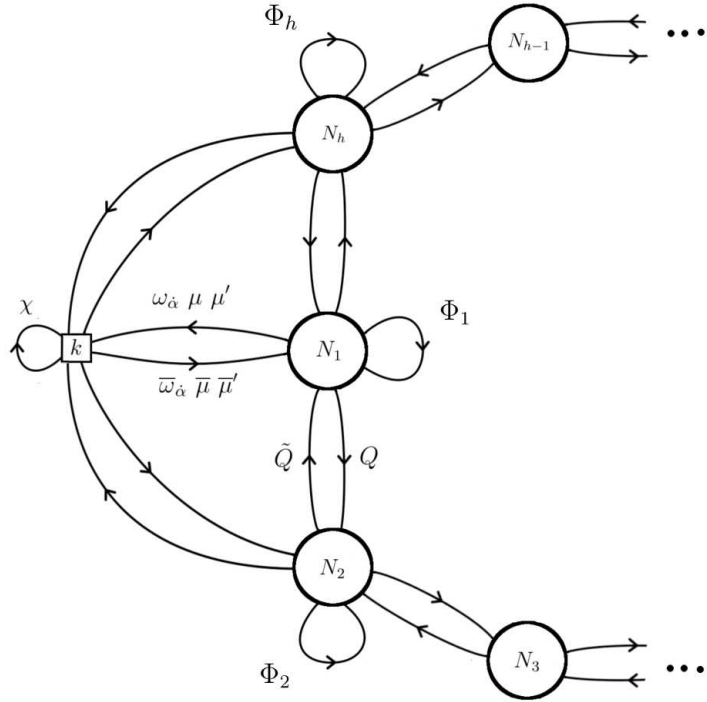}
  \caption{$\mathbb{C}^2/\mathbb{Z}_h\times \mathbb{C}$ quiver diagram with $k$ instantons placed at node $1$.}
  \label{bigquiver}
\end{figure}

\begin{figure}[ht]
  \centering
  \includegraphics[scale=.26]{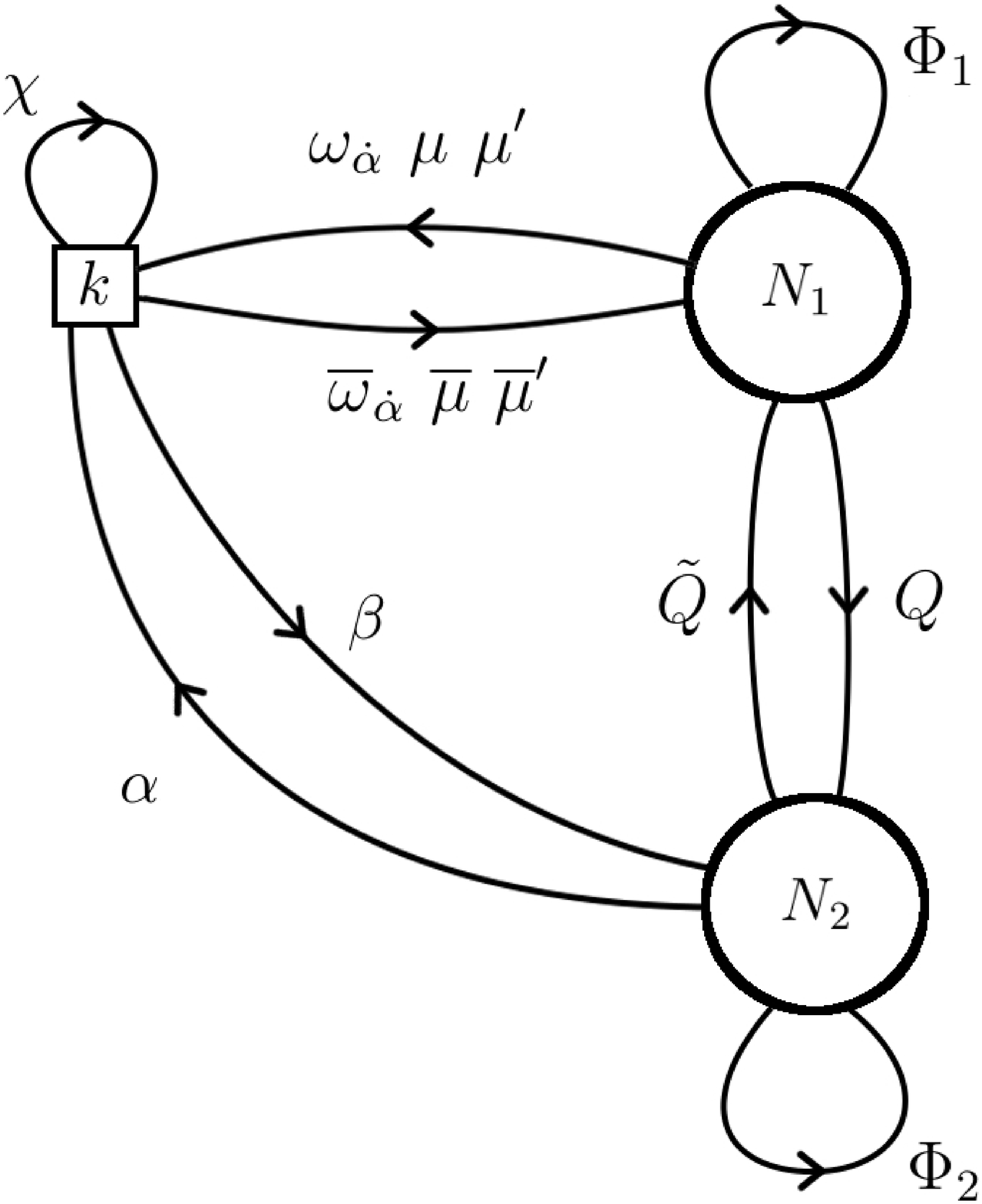}
  \caption{The ${\cal N}=2$ $U(N_1) \times U(N_2)$ quiver gauge theory, where we set $N_1=1$ and $N_2=N$ and where the bi-fundamental matter fields are 
  denoted by $Q$ and $\tilde{Q}$ and the adjoint scalars of the two nodes are denoted by $\Phi_1$ and $\Phi_2$. 
  The zero mode structure of the $k$ D($-1$)-instantons located at node 1 is also shown, see the text for details.}
  \label{quiver}
\end{figure}

Consider an ${\cal N}=2$ $U(1) \times U(N)$ gauge theory arising from fractional D$3$-branes 
at a $\mathbb{C}^2/\mathbb{Z}_h$ singularity, with $h>2$,\footnote{The calculations and results 
in this section trivially carry over to the case where $h=2$ with the only modification that, due 
to the $\mathbb{C}^2/\mathbb{Z}_2$ quiver structure, the gauge node under consideration is subject 
to a doubling of the number of flavors; for further details see Subsection \ref{comments}.}
where the fractional D$3$-brane rank assignment has been 
chosen to be $(1,N,0,\cdots,0)$. See Figure \ref{bigquiver}. We are interested in the effect of  fractional D$(-1)$-instantons with charge $k$,
located at the $U(1)$ node of the quiver. 
The spectrum of instanton zero modes, corresponding to the massless modes of the open strings with at least one endpoint 
attached to the D$(-1)$-instantons, and their interactions, captured by the moduli action, can be obtained straightforwardly 
by starting from a D$3/$D$(-1)$ system \cite{Green:2000ke,Billo:2002hm}, or a D9/D5 system \cite{Dorey:2002ik}, in flat space and performing 
an orbifold projection, see e.g.~\cite{Argurio:2008jm} for details concerning the zero mode spectrum and the orbifold projection. 
This $U(1)\times U(N)$ quiver gauge theory with the $k$ fractional D($-1$)-instantons located at the $U(1)$ node and the associated zero modes are
depicted in Figure \ref{quiver}.

The zero modes in the neutral sector, consisting of those that do not transform under the gauge group of the fractional D3-branes, arise from the open strings that have both their endpoints on the D($-1$)-instantons. This sector involves the modes $a^\mu, M^\alpha$ and  $M'^\alpha$, which are traceless $k\times k$ matrices. The traces of these modes correspond to the broken supertranslation generators and they play the role of the $\mathcal{N}=2$ superspace coordinates $x^\mu$, $\theta^\alpha$ and $\theta^{\prime \alpha}$. The neutral sector also contains the modes $\lambda^{\dot\alpha},\lambda'^{\dot\alpha}, D^c$ and $\chi$ which are (non-traceless) $k\times k$ matrices. The mass dimension and the number of Chan-Paton degrees of freedom of all the zero modes are summarized in Table 1.
\begin{table}[htdp]
\begin{center}
\label{tab_dim}
\begin{tabular}{|c|c|c|c|c|c|c|c|c|c|c|}
  \hline
 & $x$ & $\theta$ & $a$ & $\chi$ & $M$ & $\lambda$ & $D$ & $\omega$ & $\mu $ & $\alpha,\beta$  \\ \hline
 Mass dim. & -1 & -1/2 & -1 & 1 & -1/2 & 3/2 & 2 & -1 & -1/2 &  -1/2 \\
  \hline
  CP d.o.f. & 1 & 1 & $k^2$-1 & $k^2$ & $k^2$-1 & $k^2$ & $k^2$ & $k$ & $k$ & $kN$ \\
  \hline
  \end{tabular}
  \caption{The mass dimension and the number of Chan-Paton degrees of freedom of the instanton zero modes.}
\end{center}
\end{table}

The zero modes in the charged sector arise from the open strings stretching between the D($-1$)-instantons and the fractional D3-branes. 
Specifically, from the strings stretching to the single fractional D3-branes at node 1, we obtain the bosonic modes $\omega_{\dot\alpha}$ 
and the fermionic modes $\mu, \mu'$ (with the conjugate modes $\ov\omega_{\dot\alpha}, \ov\mu, \ov\mu'$) and they all have $k$ components.
Furthermore, from the strings stretching between the D($-1$)-instantons and the $N$ fractional D3-branes at node 2,
we obtain the fermionic modes $\alpha$ and $\beta$ and their Chan-Paton factors are $N\times k$ and $k \times N$ matrices, respectively.  

The 0-dimensional worldvolume moduli action for the $k$ D($-1$)-instantons for this configuration is given by
\begin{eqnarray}
\label{Smod} 
&& S_{\mathrm{moduli}}^{k}=\mathrm{tr}_k \left\{S_1+S_2+S_3+S_4\right\}  \ , \\
\mathrm{with}&&\nn \\ 
S_1 & = &  - i D^c\!\left( \ov\omega^{\dot \alpha}
(\tau^c)^{\dot\beta}_{\dot\alpha} \omega_{\dot\beta} + i
\bar\eta^c_{\mu\nu}  {[a^\mu, a^\nu]}\right)
+ i \left(\ov\mu \omega_{\dot\alpha} +
\ov\omega_{\dot\alpha} \mu + \sigma^\mu_{\beta\dot\alpha}{[M^{\beta}, a_\mu]}\right)\! \lambda^{\dot\alpha}  \nn\\
 && + i \left(\ov\mu' \omega_{\dot\alpha} + \ov\omega_{\dot\alpha} \mu' + \sigma^\mu_{\beta
\dot\alpha}{[M'^{\beta}, a_\mu]}\right)\! \lambda'^{\dot\alpha} -[a_\mu,\chi^{\dagger}] [a^\mu,\chi ]-
\frac{i}{2}M^{\alpha} [\chi^{\dagger}, M'_\alpha] \nn \\
S_2 & = & \frac{1}{2}\overline{\omega}_{\dot{\alpha}}
\omega^{\dot{\alpha}}( \chi^\dagger\chi +\chi \chi^\dagger)
+ \frac{i}{2}\left( \ov\mu' \mu \chi^\dagger - \ov\mu \mu' \chi^\dagger + \beta (\Phi_2 - \Phi_1) \alpha+\beta\alpha  \chi\right)
 \nn \\
S_3 & = & \frac{1}{2g_0^2}\left(  D_c^2 -
i  \lambda_{\dot \alpha} [\chi,
\lambda'^{\dot \alpha}] +  [ \chi,\chi^\dagger]^2 \right)\nn\\
S_4&=&  \frac{1}{2}\ov\omega_{\dot\alpha} (\widetilde{Q}^\dagger \widetilde{Q}+ Q Q^\dagger  )\omega^{\dot\alpha}
-\frac{i}{2}\left(\beta \widetilde{Q} \mu - \ov\mu Q \alpha    + \beta Q^\dagger \mu' - \ov \mu' \widetilde{Q}^\dagger \alpha \right)\nn
\end{eqnarray}
where $\tau^c$, $\eta^c_{\mu\nu}$ and $\sigma^\mu$ are the Pauli matrices, the 't~Hooft symbols and one of the off-diagonal Weyl 
blocks in the four-dimensional gamma-matrices, respectively, see \cite{Dorey:2002ik,Billo:2002hm} for their precise definitions. 
We have used the freedom to shift the modes $\chi$ by (the VEV of)
$\Phi_1=\Phi_1 {\bf 1}_{k}$. 
Note that, while the node 2 
adjoint scalar $\Phi_2$ is an $N\times N$ matrix (whose diagonal components are henceforth denoted with $\phi_{2,i}$), 
the node 1 scalar $\Phi_1$ has only one single component (henceforth denoted with $\phi_1$). 

The $S_3$-part of \eqref{Smod} contains the 0-dimensional (dimensionful) coupling constant $g_0^2 = g_s /\alpha'^2$, where $g_s$ and $\sqrt{\alpha'}$ 
are the string coupling and the string scale, respectively. In the field theory limit we send $\alpha' \to 0$ while keeping $g_s$ fixed, implying
that the action $S_3$ vanishes and the modes $D^c$, $\lambda_{\dot\alpha}$ and $\lambda'_{\dot\alpha}$ in the $S_1$-part of \eqref{Smod} 
become Lagrange multipliers enforcing the bosonic and fermionic  ADHM constraints \cite{Atiyah:1978ri}. 

From the mass dimensions of the moduli fields, given in Table~1, we see that the dimension of the measure of the corresponding moduli space 
integral is $k(N-2)$, implying the presence of a dimensionful prefactor $M_s^{k(2-N)}$, where $M_s=1/\sqrt{\alpha'}$. 
By factoring out the part of the measure involving the zero modes corresponding to the center of mass motion of the D($-1$)-instantons,  
which provide the integration measure of the chiral $\mathcal{N}=2$ superspace, we obtain the centered moduli space integral in the form 
of a contribution from the charge $k$ D($-1$)-instanton sector to the prepotential of the worldvolume gauge theory of the fractional D3-branes,
\be
\label{Fk}
     \mathcal{F}_k^s\propto M_s^{k(2-N)}e^{2\pi i k \tau_{D}} \int \mathrm{d} \{a\, M\,M'\,\lambda\,\lambda'\,D\,\chi\,\omega\,\mu\,\mu'\,\alpha\,\beta\} \mathrm{e}^{-S_\mathrm{moduli}^k}+\dots\ ,
\ee
where $-2\pi i \tau_{D}$ is the complexified instanton action for a fractional D($-1$)-instanton at the $U(1)$ node.

\subsubsection{The $k=1$ prepotential}

Let us start by explicitly evaluating the prepotential contribution \eqref{Fk} for a single $k=1$ D($-1$)-instanton at the $U(1)$
node in the quiver shown in Figure \ref{quiver}. Since we are interested in the Coulomb branch of the theory we take the VEV's of 
the hypermultiplet scalars $Q$ and $\widetilde{Q}$ to be vanishing, implying the vanishing of the $S_4$-part in \eqref{Smod}. Moreover, we take the field theory limit in which the $S_3$-part of \eqref{Smod} also vanishes. Thus, we are left with a moduli action given by the following two parts,
\begin{eqnarray}
S_1 & = &  - i D^c \ov\omega^{\dot \alpha}
(\tau^c)^{\dot\beta}_{\dot\alpha} \omega_{\dot\beta}
+ i \left(\ov\mu \omega_{\dot\alpha} +
\ov\omega_{\dot\alpha} \mu \right)\! \lambda^{\dot\alpha}  + i \left(\ov\mu' \omega_{\dot\alpha} + \ov\omega_{\dot\alpha} \mu' \right)\! \lambda'^{\dot\alpha}  \label{Sk11}\\
S_2 & = & \overline{\omega}^{\dot{\alpha}}   \omega_{\dot{\alpha}} \overline{\chi}\chi
+ \frac{i}{2}\left( \ov\mu' \mu \ov\chi - \ov\mu \mu' \ov\chi + \beta (\Phi_2 - \Phi_1+\chi) \alpha \right) ~.\label{Sk1}
\end{eqnarray}
Since the $\lambda$-modes appear linearly in the moduli action we begin by integrating over them, which brings down a factor, 
\begin{eqnarray}
   \mu \mu' \overline{\mu} \overline{\mu}' 
   \left(\overline{\omega}^{\dot\alpha} \omega_{\dot\alpha} \right)^2~.
\end{eqnarray}
This factor is used when integrating out all the 4 modes $\mu,\mu',\overline{\mu}$ and $\overline{\mu}' $. 
Therefore,  in order to get a non-vanishing contribution to the prepotential,  
the second and third term in \eqref{Sk1} should be expanded only to zeroth order and hence we discard them. 
The integrals over the charged fermion zero modes $\alpha$ and $\beta$ make use of the last term in \eqref{Sk1} and bring down a determinant,
\begin{eqnarray}
\label{det}
 \det \left( \Phi_{2} -\Phi_1 {\bf 1}_{N} +\chi {\bf 1}_{N}\right) ~.
\end{eqnarray} 
Since this determinant depends holomorphically on $\chi$ (and not on $\overline{\chi}$), and since there is no term in the remaining moduli action that can be brought down in order to compensate for the complex phase of the  $\chi$'s, the integration over $\chi$ and $\overline{\chi}$ will only give a non-vanishing result for terms arising from the determinant \eqref{det} which do not contain any power of $\chi$. Hence, we can remove $\chi$ from \eqref{det} and perform the Gaussian integrals over $\chi$ and $\overline{\chi}$, which yield an inverse factor of $\overline{\omega}^{\dot\alpha} \omega_{\dot\alpha}$.

By combining these results we conclude that the $k=1$ stringy instanton contribution to the prepotential has the following form,
\be
\label{Fk1}
     \mathcal{F}_{1}^s= \mathcal{N}_1 
     M_s^{2-N}e^{2\pi i  \tau_{D}} \det \left(\Phi_{2}- \Phi_1 {\bf 1}_{N}\right)\times \mathcal{I}
\ee
where $\mathcal{N}_1$ is an overall normalization constant and the remaining bosonic integral is given by
\begin{equation}
\label{I}
\mathcal{I} =\int  d^2\omega_{\dot\alpha} d^2 \ov\omega^{\dot\alpha}d^3D^c~  \overline{\omega}^{\dot{\alpha}} \omega_{\dot{\alpha}} \,
e^{ -i D^c\, \overline{\omega}^{\dot{\alpha}} (\tau^c)_{\dot{\alpha}}^{ \dot{\beta}} \omega_{\dot{\beta}} }~.
\end{equation}
The integral $\mathcal{I}$ is a dimensionless number and our remaining task is to show that this number is non-zero. As the integral  \eqref{I} stands, it is not well-defined and therefore we are required to regularize it. We will evaluate \eqref{I} using 3 different regularization methods. The key feature of all these regularization methods is to introduce a dimensionful parameter which keeps the instanton from shrinking to zero size by smoothing out the corresponding singularity in the instanton moduli space.

In the first regularization we allow the hypermultiplet scalars to
acquire a non-vanishing VEV, such that $\widetilde{Q}^\dagger
\widetilde{Q}+ Q Q^\dagger =|v|^2\neq 0$.\footnote{See, for example,
  \cite{Petersson:2007sc} for an $\mathcal{N}=1$ configuration in
  which this regularization was used.} This implies that the
$S_4$-part in \eqref{Smod} is now non-vanishing and should be added to
\eqref{Sk1}. Note that all the terms in the $S_4$-part in \eqref{Smod}
that involve fermion zero modes are irrelevant in this configuration
since they contain one of the modes $\mu,\mu',\overline{\mu}$ or
$\overline{\mu}' $, which have already been soaked up. By taking this non-vanishing VEV into account, the integral \eqref{I} becomes well-defined and gives the following finite result,
\begin{equation}
\label{Iv}
\mathcal{I}_v =\int  d^2\omega_{\dot\alpha} d^2 \ov\omega^{\dot\alpha}d^3D^c~  \overline{\omega}^{\dot\alpha}  \omega_{\dot\alpha} \,
   e^{ -i  D^c\, \overline{\omega}^{\dot{\alpha}} (\tau^c)_{\dot{\alpha}}^{\dot{\beta}} \omega_{\dot{\beta}}
   -\frac{1}{2} |v|^2 \overline{\omega}^{\dot{\alpha}}  \omega_{\dot{\alpha}}}= 8\pi^4
\end{equation}
where the explicit steps of the evaluation are given in Appendix \ref{vevreg}. As expected,  the result in \eqref{Iv} is independent of the VEV $v$ we deformed the integral \eqref{I} with.  

A common way to regularize instanton partition functions is to consider the gauge theory to be defined on a non-commutative space. In such theories, the ADHM constraints are deformed by the presence of a non-commutativity parameter $\xi$.  One consequence of this deformation is that even gauge theories with an Abelian $U(1)$ 
gauge group allow for non-trivial instanton configurations \cite{Nekrasov:1998ss}. In terms of the instanton moduli action, the non-commutative deformation amounts to simply adding a term $i D^c\xi\delta_{c3}$ to \eqref{Sk11}, where $\delta_{c3}$ is a Kronecker delta. This procedure renders the integral \eqref{I} well-defined and we obtain
 \begin{equation}
\label{Ixi}
\mathcal{I}_\xi =\int  d^2\omega_{\dot\alpha} d^2 \ov\omega^{\dot\alpha}d^3D^c~  \ov\omega^{\dot\alpha}\omega_{\dot\alpha}\,e^{ -i D^c \ov\omega^{\dot \alpha}
(\tau^c)^{\dot\beta}_{\dot\alpha} \omega_{\dot\beta} +i D^c\xi\delta_{c3} }= 8\pi^4
\end{equation}  
where the intermediate steps are provided in Appendix \ref{nonreg}. 

In the third way of regularizing, we refrain from taking the field theory limit by keeping $\alpha'$, and hence $g_0$, finite.\footnote{This regularization was used, for instance, in \cite{Green:2000ke} for an $\mathcal{N}=4$ configuration and in \cite{Ferretti:2009tz} for an $\mathcal{N}=1$ configuration.} This implies that the $S_3$-part of \eqref{Smod}, which for $k=1$ reduces to only the first term in $S_3$, no longer vanishes. By adding the term $D_c^2/(2g_0^2)$ to \eqref{Sk11}, the integral \eqref{I} is again well-defined and gives the result,
    \begin{equation}
\label{Ial}
\mathcal{I}_{\alpha^\prime}=  \int  d^2\omega_{\dot\alpha} d^2 \ov\omega^{\dot\alpha}d^3D^c~  \ov\omega^{\dot\alpha}\omega_{\dot\alpha}\,e^{ -i D^c \ov\omega^{\dot \alpha}
(\tau^c)^{\dot\beta}_{\dot\alpha} \omega_{\dot\beta} -\frac{1}{2g_{0}^{2}}  (D^c)^2}= 8\pi^4
\end{equation}   
where the details are given in Appendix \ref{areg}. 
 
We conclude that the 3 regularization methods indeed yield the same non-vanishing number, $\mathcal{I}_v=\mathcal{I}_\xi=\mathcal{I}_{\alpha^\prime}=8\pi^4$ which can be absorbed in the string scale in \eqref{Fk1}. The fact that these regularized integrals are independent of their corresponding dimensionful parameter is simply a consequence of the invariance of \eqref{I} under rescalings in terms of a parameter $k$, e.g.~$D^c\to k^2D^c$, $\omega\to(1/k)\,\omega$ and $\bar\omega\to(1/k)\,\bar\omega$. Hence, the dimensionful parameters are rescaled away by choosing $k=v$ in \eqref{Iv}, $k=1/\sqrt{\xi}$ in \eqref{Ixi}  and $k=g_0$ in \eqref{Ial}.

\subsubsection{The $k>1$ prepotentials}

We have seen that a D($-1$)-instanton on the $U(1)$ node of the quiver gauge theory in Figure \ref{quiver} 
gives a non-vanishing contribution to the prepotential of the form displayed in \eqref{Fk1}. 
In particular, we saw that using a finite VEV for the hypermultiplet scalars or a finite value of $\alpha'$ 
was equivalent to using the non-commutative regularization. The latter regularization is the one employed, 
for example, by Nekrasov in \cite{Nekrasov:2002qd} when deriving the instanton contributions to the prepotential in a 
$U(N_c)$ $\mathcal{N}=2$ gauge theory with $N_f$ flavors.\footnote{As
  usual in D-brane set ups, the gauge group is really $U(N_c)$ rather than
  $SU(N_c)$, though the Abelian factor is usually discarded. 
Note that the construction of
  \cite{Nekrasov:2002qd} is also based on a $U(N_c)$ gauge group.}
This equivalence among regularizations motivates us to relate the
stringy D($-1$)-instanton contribution in \eqref{Fk1} to the Nekrasov
instanton formulae, in the same way as we noted in the $Sp(0)$ case that
the stringy instanton expressions were completely equivalent to the
continuation to $N_c=0$ of the $Sp(N_c)$ expressions obtained in gauge
theory.
 
Let us observe first that the Nekrasov formulae are for gauge instanton contributions in a non-Abelian $U(N_c)$ theory
where $N_c>1$. We thus assume that the stringy instanton corrections
are just given by the formulae obtained in  \cite{Nekrasov:2002qd} extended to the Abelian case where $N_c=1$ and $N_f=N$. The ``stringy'' character of such contributions
consists in having instanton corrections in an Abelian (commutative)
theory.

Further motivation to the extension of Nekrasov's formulae to the Abelian $N_c=1$ case as a mean to obtain
stringy instanton results can be based on the analysis of the associated brane constructions.
Indeed, $k$ D($-1$)-instantons lying on top of a single D$3$-brane yield an instantonic configuration that,
although stringy from the field theory viewpoint, 
possesses the same moduli structure as an ordinary instanton.

The generic form of the prepotential in the $U(N_c)$ $\mathcal{N}=2$ gauge theory is given by \eqref{FTinst} but where $\Lambda$ now is the strong coupling scale
of the $U(N_c)$ gauge group and the $1$-loop coefficient of the $\beta$-function is given by $b_0=2N_c-N_f$. 
The instanton corrections to the prepotential up to $k=3$ can be expressed as follows \cite{Nekrasov:2002qd},
\begin{equation}\label{f1}
 \begin{split}
 F_1^g =& \sum_{u=1}^{N_c} S_u \\
 F_2^g =& \sum_{u\neq v}^{N_c} \frac{S_u S_v}{\phi_{uv}^2} 
       +\frac{1}{4} \sum_{u=1}^{N_c} S_u \frac{\partial^2 S_u}{\partial \phi_u^2}\\
 F_3^g =& \frac{1}{36} \sum_{u=1}^{N_c} S_u \Bigg[ S_u \frac{\partial^4 S_u }{\partial \phi_u^4}     
                + 2 \frac{\partial S_u }{\partial \phi_u} \frac{\partial^3 S_u }{\partial \phi_u^3} 
                + 3 \frac{\partial^2 S_u }{\partial \phi_u^2} \frac{\partial^2 S_u }{\partial \phi_u^2} \Bigg] +\\
             &  + \sum_{u\neq v}^{N_c}  \frac{S_u S_v}{\phi_{uv}^4} \Bigg[5 S_u 
                - 2 \phi_{uv} \frac{\partial S_u }{\partial \phi_u}  
                + \phi_{uv}^2 \frac{\partial^2 S_u }{\partial \phi_u^2}\Bigg]\\
             &  + \sum_{u\neq v,v\neq w,w \neq u}^{N_c}  \frac{2 S_u S_v S_w}{3 \phi_{uv}^2
             \phi_{vw}^2 \phi_{uw}^2} \Big[ \phi_{uv}^2 + \phi_{vw}^2 + \phi_{wu}^2 \Big]
 \end{split}                
\end{equation}
where $\phi_u$ indicates the $u$-th diagonal component of the adjoint field $\phi$ and $\phi_{uv} = \phi_u - \phi_v$. Moreover, the function $S_u=S_u(\phi_u)$ for a unitary gauge group is
given by,
\begin{equation}\label{S}
 S_u(\phi_u) = \frac{\prod_{j=1}^{N_f} (\phi_u + m_j)}{\prod_{v\neq u}^{N_c} \phi_{uv}^2}~.
\end{equation}
In order to compare first the $k=1$ prepotential in \eqref{f1} with the D($-1$)-instanton stringy contribution in \eqref{Fk1}, 
we have to set $N_c=1$ and $N_f=N$, such that $\phi_u\to\phi$ and $S_u\to S = \prod_{j=1}^{N} (\phi + m_j)$. 
Moreover, by mapping $\phi\to -\phi_1$  and $m_j\to \phi_{2,j}$ we obtain, 
\be
\label{FTFk1}
\mathcal{F}_1^s= \Lambda^{ b_0} \prod_{j=1}^{N} (  \phi_{2,j}-\phi_1)
\ee
which indeed has the same form as \eqref{Fk1}
(when we diagonalize the matrix $\left( \Phi_{2}- \Phi_1 {\bf 1}_{N}\right)$ in \eqref{Fk1} and 
express it in terms of its eigenvalues);
we also identify $M_s^{2-N} e^{2 \pi i \tau_D} \equiv \Lambda^{b_0}$, where
 $b_0=2-N$ and we have taken ${\cal N}_1 {\cal I}=1$.
Note that this is not the 1-loop beta function coefficient for an
Abelian gauge theory, hence $\Lambda$ above is {\em not} the scale of
the Landau pole of this IR free theory, but an independent scale.  
However, it is precisely the 1-loop beta function coefficient one would expect for a non-commutative $\mathcal{N}=2$ 
Abelian gauge theory \cite{Martin:1999aq,Khoze:2000sy}, in agreement with the fact that the formul\ae\ in \cite{Nekrasov:2002qd}, 
from which we extracted \eqref{FTFk1}, were obtained for a non-commutative $U(N_c)$ gauge theory. 

To get a feeling that one indeed obtains a non-vanishing result for
\eqref{Fk} for a general value of $k$, we now
perform a simple fermion zero mode counting, in order to check that the D($-1$)-instanton moduli space integral 
allows us to integrate out all the fermion zero modes. As before, we begin by considering the Coulomb branch in the field theory limit, 
i.e.~when $S_3$ and $S_4$ in \eqref{Smod} are vanishing. In this limit, integrating over the modes $\lambda$ and 
$\lambda'$ gives rise to the $4 k^2$ fermionic ADHM constraints, 
\be
    \prod_{\dot\alpha,i,j}(\ov\mu \omega_{\dot\alpha} +
    \ov\omega_{\dot\alpha} \mu + \sigma^\mu_{\alpha \dot\alpha}{[M^\alpha, a_\mu]})_i^j
   \prod_{\dot\beta,m,n} (\ov\mu' \omega_{\dot\beta} + \ov\omega_{\dot\beta} \mu' +
    \sigma^\mu_{\beta \dot\beta}{[M'^\beta, a_\mu]})_m^n 
    \label{fermiADHM}
\ee
with $i,j,m,n=1\dots k$. Since there are $4k^2-4$ of the $M,M'$ modes and $4k$ of the $\mu,\mu',\ov{\mu},\ov{\mu}'$ modes,
after using the $4 k^2$ constraints, there will remain $4k-4$ modes of the set $\{M,M',\mu,\mu',\ov{\mu},\ov{\mu}' \}$. 
We can use the terms $M^{\alpha} [\chi^{\dagger}, M'_\alpha] $, $\ov\mu' \mu \chi^\dagger$ and $\ov\mu \mu' \chi^\dagger$ in 
the $S_1$ and $S_2$-parts of \eqref{Smod} in order to integrate out these remaining $4k-4$ modes in this set,
bringing down a factor of $\left(\chi^\dagger \right)^{2k-2}$. In order to cancel the complex phase of this factor, 
we can use the term  $\beta\alpha \chi$ in the $S_2$-part of \eqref{Smod} and hence $4k-4$ modes of the set $\{\alpha,\beta\}$. 
Since there are $2Nk$ of this latter set, there will still remain $2Nk-4k+4$ of these modes. 
Under the simplifying assumption that all the fluctuations $\phi_{2,j}$ are equal to a single 
field $\phi_2$, if we use the term $\beta (\Phi_2 - \Phi_1) \alpha$ in
the $S_2$ part in order 
to integrate out these remaining modes, we bring down a factor,
\be 
\label{factor}
(\phi_2 - \phi_1)^{Nk-2k+2}=\frac{(\phi_2 - \phi_1)^{Nk}}{(\phi_2 - \phi_1)^{2k-2}}
\ee
where, in the numerator of the RHS, we see the function in \eqref{S}
for $N_c=1$ and $N_f=N$, 
\be\label{sinst}
S(\phi)=\prod_{j=1}^{N} (\phi + m_j)\to (\phi_2-\phi_1)^N\ , 
\ee
appearing to the power $k$, in agreement with $k=1,2,3$ prepotentials in \eqref{f1}. Note that the dimension of \eqref{factor} is in agreement with the dimensionful prefactor $M_s^{k(2-N)}$ of the centered moduli space measure in \eqref{Fk}. 

In summary, we have argued that the evaluation of the stringy D($-1$)-instanton moduli space integral \eqref{Fk} 
is equivalent to the Nekrasov evaluation \cite{Nekrasov:2002qd}, or even identical if we use the non-commutative regularization. 
We thus report the explicit stringy results up to $k=3$ computed by means of Nekrasov formulae extended to
$N_c=1$; 
this amounts to performing the following substitutions in \eqref{f1}, 
\begin{equation}
 \phi \rightarrow -\phi_1\ , \ \ \ \ \ \ \
 m_j  \rightarrow \phi_{2,j}\ .
\end{equation}
Eventually, the stringy contributions to the prepotential for $k=1,2,3$ are
\begin{eqnarray}\label{1sUgen}
\mathcal{F}_1^s &=&  \Lambda^{ b_0}\ \det ( \Phi_2- \Phi_1)\ , \\ \label{2sUgen}
\mathcal{F}_2^s &=&  \frac{\Lambda^{2 b_0}}{2}\det( \Phi_2- \Phi_1) e_{N-2}( \Phi_2- \Phi_1)\ , \\ \label{3sUgen}
\mathcal{F}_3^s &=&  \frac{\Lambda^{3 b_0}}{3}\det ( \Phi_2- \Phi_1)\Big(2\det( \Phi_2- \Phi_1)e_{N-4}( \Phi_2- \Phi_1) \\
&& \quad+e_{N-1}( \Phi_2- \Phi_1)e_{N-3}( \Phi_2- \Phi_1)+\big(e_{N-2}( \Phi_2- \Phi_1)\big)^2\Big)\ , \nonumber
\end{eqnarray}
where the $e_i$ where defined in \eqref{edef}.
Considering again all the fluctuations $\phi_{2,j}$ to be equal to $\phi_2$ we have the simpler expressions
\begin{eqnarray}\label{1sU}
\mathcal{F}_1^s &=&  \Lambda^{ b_0}\ ( \phi_2- \phi_1)^N\ , \\ \label{2sU}
\mathcal{F}_2^s &=&  \frac{\Lambda^{2 b_0}}{4}\ N (N-1)\ ( \phi_2- \phi_1)^{2N-2}\ , \\ \label{3sU}
\mathcal{F}_3^s &=&  \frac{\Lambda^{3 b_0}}{6}\ N (N-1)^3\ ( \phi_2- \phi_1)^{3N-4}\ .
\end{eqnarray}
In the following Subsections we are going to compare these stringy instanton contributions
against the corrections introduced by ordinary gauge instantons in a different gauge theory
representing a specific UV completion of the ${\cal N}=2$ $U(1) \times U(N)$ model at hand.

\subsection{UV and IR unitary theories}
\label{Uflow}

In order to interpret the stringy instantons in terms of ordinary gauge 
instantons, we must complete the $U(1) \times U(N)$ theory in the ultraviolet. 
The $U(1)$ gauge group can be naturally
interpreted as resulting from a  Higgsing of a larger gauge group $U(N_c)$.
The Higgsing procedure corresponds to considering a point in the moduli 
space where the scalar field $\Phi_1$ transforming in the adjoint representation of the gauge group
acquires a non-trivial VEV.
In general, a complete breaking of the original $U(N_c)$ gauge group 
leads to a $U(1)^{N_c}$ low-energy theory. 
In order to be able to neglect $N_c-1$ out of the $N_c$ $U(1)$ factors of the low-energy theory, 
we need to have a Higgs branch for each of them.
This can be achieved by identifying a specific point in the moduli space for $\Phi_1$
and assigning specific masses to the flavors \cite{Argyres:1996eh}.

To be concrete, the UV theory we consider is still described by 
the $\mathbb{C}^2/\mathbb{Z}_h$ quiver diagram introduced in the previous Subsection, 
but now the rank assignment is
\begin{equation}\label{ranks}
 (P+1\ ,\ N\ ,0,\ 0\ ,...,\ P\ )_h \,
\end{equation}
and then the symmetry group is given by
\begin{equation}\label{sym_stru}
  \underbrace{U(P+1)}_{\text{color}} \times \underbrace{U(N) \times U(P)}_{\text{flavor}}\ ,
\end{equation}
where we have introduced one positive parameter $P$;
the UV completion \eqref{sym_stru} can be thought of as resulting from the addition 
of $P$ fractional branes on nodes $1$ and $h$ of the quiver.
For later convenience, we also define
\begin{equation}\label{cf}
 N_c = P+1\ , \ N_f = P + N\ ,
\end{equation}
which represent the number of ``colors'' and ``flavors'' of the UV theory.

In the moduli space of the UV theory, we consider the specific point where
the adjoint field $\Phi_1$ breaks completely the gauge symmetry by acquiring
the following non-degenerate VEV 
\be
\label{phi1vev}
\Phi_1 = (-x_1 M, \dots,-x_{N_c-1}M, -\phi_1)_{N_c}\ ,
\ee
where $x_i$ are $N_c-1$ different numbers of order one which break the gauge
group $U(N_c)$ down to $U(1)^{N_c-1}\times U(1)$, and $\phi_1$ is a field fluctuation 
associated to the last $U(1)$ factor.
We also add a diagonal mass term $m$ for the $N_f$ flavors.
It can be explicitly checked that, in order to have a Higgs branch 
at low energy for each of the $U(1)^{N_c-1}$ factors,
we have to assign the non-trivial masses 
\be
\label{massa}
 m=(x_1 M, \dots, x_{N_c-1} M,\underbrace{0,\dots, 0}_N )_{N_f}\ ;
\ee
the unbroken, low-energy flavor group is therefore $U(N)$.

The mass assignment \eqref{massa} can be interpreted in terms of a specific non-vanishing VEV
for the adjoint field $\Phi_h$ associated to the $h$-th node of the quiver, namely
\begin{eqnarray}\label{phi3vev}
 \Phi_h = (x_1 M, \dots, x_{N_c-1} M )_P\ . 
\end{eqnarray}
The field $\Phi_2$, which can be associated to the last $N$ components in the mass array \eqref{massa},
does not acquire any VEV; we parametrize its components as follows:
\begin{equation}\label{phi2vev}
 \Phi_2 = (\phi_{2,1}, ..., \phi_{2,N})_N\ .
\end{equation}
Taking these fluctuations into account in the mass array we obtain
\be
\label{massa_flu}
 m=(x_1 M, \dots, x_{N_c-1} M ,\phi_{2,1}, ..., \phi_{2,N})_{N_f}\ .
\ee
In the next Subsections we will consider all $\phi_{2,i}$ to be equal and we will consequently
drop the label $i$.

To summarize, the $U(1) \times U(N)$ IR theory is re-obtained by considering the large mass limit $M\rightarrow \infty$
in the UV theory \eqref{sym_stru} at the point of the moduli space specified by
(\ref{phi1vev}), (\ref{phi3vev}) and \eqref{phi2vev}.

\subsection{Comparing gauge and stringy instantons}

In this Subsection we show that the gauge instanton computation exactly reproduces the stringy instanton results we have presented in Subsection \ref{Uinst}, at least up to three instantons.

Recalling Eq.~\eqref{cf} relating $N$ and $P$ (introduced in \eqref{ranks}) 
to $N_c$ and $N_f$, the one-loop $\beta$-function coefficient for the UV theory is
\begin{equation}\label{b1}
b_0= 2N_{c} - N_{{f}} = P - N + 2 \ .
\end{equation}
We require asymptotic freedom for the UV theory, namely $b_0>0$.
In this regime, in order to compute the instanton contributions to the prepotential of 
the UV theory, we can use ordinary instanton techniques.
In particular, the ordinary gauge instanton corrections to the prepotential of an 
asymptotically free ${\cal N}=2$, $SU(N_{c})$ gauge theory with $N_{f}$ flavors have been explicitly 
computed in \cite{Nekrasov:2002qd} using localization techniques, 
and for instance in \cite{D'Hoker:1996nv} using the Seiberg-Witten curve. 

\subsubsection{$1$-instanton contribution}

In order to obtain the prepotential instanton contribution on the moduli space point
that we have identified in the previous section, we first evaluate the $S_k(\phi_k)$
function \eqref{S} on that point.
We replace $\phi_k$ with the components of $\Phi_1$ in \eqref{phi1vev}
and the mass array as in \eqref{massa_flu} with all
$\phi_{2,i} = \phi_2$. The function $S_k(\phi_k)$ simplifies to
\begin{equation}
 S_k(\phi_k) = \delta_{k,N_c} \frac{ (\phi_2 - \phi_1)^N \prod_{j=1}^{N_c-1} (  x_j M-\phi_1)}
 {\prod_{l=1}^{N_c-1} ( x_l M-\phi_1)^2}\ ,
\end{equation}
where, according to the VEV assignment \eqref{phi1vev}, we have $\phi_{N_c} = - \phi_1$.
In the limit of large mass $M$, the leading term of $S_k(\phi_k)$ is
\begin{equation}\label{sgauge}
 S_k(\phi_k) = \delta_{k,N_c}\ \frac{ (\phi_2 - \phi_1)^N}
 {\text{det}\ M} + {\cal O} \left(M^{-N_c} \right)
\end{equation}
where
\begin{equation}
 \text{det}\ M =  \prod_{l=1}^{P} x_l M\ .
\end{equation}
Recalling Eq. \eqref{f1}, the $1$-instanton contribution to the prepotential is
\begin{equation}\label{F1-gau}
\mathcal{F}_1^g= \Lambda^{b_0} F_1^g=
 \Lambda^{b_0} \sum_{k=1}^{N_c} S_k(\phi_k)= 
 \Lambda^{b_0} \frac{ (\phi_2 - \phi_1)^N}
 {\text{det}\ M}  \ .
\end{equation}
Eventually we have
\begin{equation}
 {\cal F}_1^s = {\cal F}_1^g\ ,
\end{equation}
or, said otherwise, the gauge instanton contribution \eqref{F1-gau} 
coincides with the stringy instanton contribution \eqref{1sU}
if we consider the matching of scales between the UV, the IR theory
and the string theory parameters as
\be
\label{matchU}
M_s^{2-N} e^{2 \pi i \tau_D}=
\Lambda_{\text{IR}}^{b_0}=\frac{\Lambda_{\text{UV}}^{b_0}}{ {\text{det}\ M}  }\ .
\ee
This accounts for the presence of the factor ${\text{det}\ M}  $ in the expression \eqref{F1-gau} for the prepotential
in the gauge theory computation.

\subsubsection{$2$-instanton contribution}

The $k=2$ gauge instanton contribution to the prepotential is given in \eqref{f1} and,
if we consider the masses \eqref{massa_flu} (with all the fluctuations equal 
to $\phi_2$) and the VEV's \eqref{phi1vev},
the first sum is vanishing; this follows from the combination of the facts 
that the sum runs on $u\neq v$ and that $S_k(\phi_k) \propto \delta_{N_c,k}$. 
We have then
\begin{equation}\label{2_term}
 F_2^g =  \frac{1}{4}\ S_{N_c}(\phi_{N_c}) 
 \ \frac{\partial^2 }{\partial \phi_k^2} S_{N_c}(\phi_k) \Bigg|_{\phi_k = \phi_{N_c}}\ .
\end{equation}
The $n$-th derivative of $S_{N_c}(\phi_k)$ for $n\leq N$ evaluated at $\phi_k=\phi_{N_c}$ is
given by
\begin{equation}\label{nth}
 \frac{\partial^{n} }{\partial X^n} S_{N_c}(\phi_k) \Bigg|_{\phi_k = \phi_{N_c}}
 = \frac{N!}{(N-n)!}\frac{(\phi_2 - \phi_1)^{N-n}}{\text{det}M} + {\cal O}\left( M^{-N_c} \right)\ .
\end{equation}
Using this general formula in \eqref{2_term}, the $k=2$ gauge instanton prepotential becomes
\begin{equation}\label{root}
 F_2^g = \frac{1}{4}\, N(N-1) \frac{ (\phi_2 - \phi_1)^{2N-2}}
 {\text{det} M^2} + {\cal O}\left( M^{-2N_c+1} \right)\ ,
\end{equation}
Once again this corresponds to the stringy instanton contribution at level $k=2$ if we 
use the matching of scale \eqref{matchU}:
\begin{equation}
 {\cal F}_2^s = {\cal F}_2^g\ .
\end{equation}

\subsubsection{$3$-instanton contribution}

We consider the $3$-instanton contribution 
to the prepotential given in \eqref{f1},
with masses \eqref{massa_flu} (again with all fluctuations equal to $\phi_2$)
and VEV's \eqref{phi1vev}.
As already happened at level $k=2$, the delta function 
structure of the function $S_u(\phi_u)$ simplifies the expressions
and the $k=3$ prepotential \eqref{f1} can be rewritten 
as follows:
\begin{equation}
 \begin{split}
   F_3^g =& \frac{1}{36} \sum_{u=1}^{N_c} S_u \Bigg[S_u \frac{\partial^4 S_u}{\partial \phi_u^4} + 2 \frac{\partial S_u }{\partial \phi_u} \frac{\partial^3 S_u }{\partial \phi_u^3} 
                + 3 \frac{\partial^2 S_u }{\partial \phi_u^2} \frac{\partial^2 S_u }{\partial \phi_u^2} \Bigg] \ ,             
 \end{split}
\end{equation}
where the symbol $S_u$ is a compact way to indicate $S_u(\phi_u)$.
Recalling the expression \eqref{nth} for the $n$-th
derivative of the function $S_{N_c}(\phi_k)$, we have
\begin{equation}
 \begin{split}
 F_3^g = \frac{N(N-1)^3}{6} \ \frac{(\phi_2 - \phi_1)^{3N-4}}{\text{det}M^3} 
  + {\cal O} \left( M^{- 3N_c + 2}\right)
  \end{split}
\end{equation}
which once again is equivalent to the stringy instanton contribution
by using the matching of scale \eqref{matchU}, namely
\begin{equation}
 {\cal F}_3^s = {\cal F}_3^g\ .
\end{equation}

At this point we hope to have provided convincing evidence of the matching between the gauge and stringy instanton computations. Actually, there is a simple argument that suggests that the matching can be extended to any order in the instanton number. Indeed, the expression for $S_k(\phi_k)$ given in \eqref{sgauge} reduces, at leading order when $M$ is large, to the expression \eqref{sinst} for $S$ in the $N_c=1$, ``stringy" case, up to the $\text{det}M$ factor. The expression from which any prepotential contribution is computed is then exactly the same, and so are the results.

\subsection{UV and IR unitary theories through duality cascade}
\label{cascadeD}

In the previous Sections we have seen that computations of the gauge instanton and stringy instanton 
contributions to the prepotential agree, order by order in the instanton number, at least up to three instantons.
In this Subsection we would like to give a slightly broader picture of the equivalence of the two computations. 

It is a well know fact that the prepotential $\mathcal{F}$ of an $\mathcal{N}=2$ theory is a holomorphic
function of the fields and of its parameters. For this reason, $\mathcal{F}$ is supposed not to jump when 
we move inside the moduli space of the theory, even if, in different points of the moduli space, 
the microscopic origin of the effective prepotential could be different. 
The continuity of the prepotential is the underlying mechanism that allows the reinterpretation of the stringy 
instantons as gauge theory instantons. Indeed, the flows we have used until now to pass from the UV theory, 
where we can do gauge theory instanton computations, to the IR theory, where we are required to do stringy 
instanton computations, imply sending some vacuum expectation values $M$ to large values,
and the prepotential should be continuous in this process.
The flows we use in Subsections \ref{Spflow} and \ref{Uflow} break completely the non-Abelian gauge group 
and change the nature of the D-brane instantons; these are interpreted as gauge theory instantons 
in the UV theory and as stringy instantons in the IR theory. However, due to the holomorphicity of the prepotential, 
the two computations should give the same results, which they do.

The flows that we have considered until now, assuming $M\gg \Lambda$, do not pass through any strongly coupled
region and hence they allow us to map gauge and stringy instantonic computations order by order.
However, the quiver gauge theories we are considering are the same ones that appear in the far IR of
the field theory side of many examples of AdS/CFT correspondence. This embedding of the field theory 
inside AdS/CFT correspondence provides a privileged flow, called the duality cascade, that can be deduced
from the dual supergravity solutions. This is a well known phenomenon in $\mathcal{N}=1$ field theories where the flow takes 
place intrinsically at strong coupling and it can be thought of as proceeding
through a series of Seiberg dualities \cite{Klebanov:2000hb}. The continuity of the superpotential along 
this flow was nicely used in \cite{GarciaEtxebarria:2007zv} to infer the equivalence of the superpotential contribution 
among theories related by Seiberg dualities; in addition, in the particular case of the last step in the cascade,
\cite{GarciaEtxebarria:2007zv} shows the equivalence 
between the gauge theory non-perturbative contribution to the superpotential of the theory before the last step of the cascade
and the stringy instantonic contribution to the superpotential of the theory after the last step of the cascade. 

In the specific cases of unitary groups, like the one considered in this Section, it is known that a similar 
cascade occurs also for $\mathcal{N}=2$ theories \cite{Benini:2008ir,Cremonesi:2009hq}. 
In the ${\cal N}=2$ case, the flow proceeds through a 
series of baryonic root transitions \cite{Argyres:1996eh}, and it reproduces the flow described by the supergravity solution. 
The baryonic root transition is a strong coupling effect of some $\mathcal{N}=2$ theories that reproduces the same 
numerology of the Seiberg duality of $\mathcal{N}=1$ theories. Namely, $\mathcal{N}=2$ SQCD with $SU(N_c)$ gauge group and $N_f$ 
fundamental hypermultiplets has, when asymptotically free (i.e. $2N_c-N_f > 0$), a quantum modification of the moduli space that splits the intersection 
of the Higgs and Coulomb branches of the theory.
At the baryonic root, namely the point in the quantum moduli space where the baryonic branch (i.e. the branch where the gauge group is fully Higgsed) intersects the Coulomb branch,
quantum effects force the adjoint field $\Phi$ in the vector multiplet to acquire the expectation value: 
 \begin{equation}
 \label{Broot}
\Phi = \Lambda (0,...,0,\omega,...,\omega^{2N_c - N_f}) \ ,
 \end{equation}
where $\Lambda$ is the strong coupling scale of the theory and $\omega$ is the $(2N_c-N_f)$-th root of unity. 
At this specific point of the moduli space, the effective theory is an $\mathcal{N}=2$ SQCD with $SU(N_f - N_c) \times U(1)^{2N_c - N_f}$
gauge group, $N_f$ fundamental hypermultiplets and $2N_c - N_f$ hypermultiplets which are singlets under the non-Abelian gauge group, 
but each one is charged under one of the $U(1)$ factors. Let us observe that in this case the UV theory is UV free, while the IR theory is IR 
free and the numerology of the rank of the color groups and the number of flavors is exactly the same as for Seiberg duality in $\mathcal{N}=1$ SQCD with $SU(N_c)$
gauge group and $N_f$ fundamental flavors. Note that the VEVs \eqref{Broot} are due to the strongly coupled dynamics of the theory. A flow at the baryonic root point of the moduli space, or close to it, is hence intrinsically at strong coupling.

The embedding in a duality cascade of the $U(1) \times U(N)$ theory which we have considered in Subsection \ref{Uinst} can be done, for example, in the following way:
consider a set of $(2N, N, 2N-1)$ fractional branes at the tip of the $\mathbb{C}^2/\mathbb{Z}_3 \times \mathbb{C}$ singularity, and perform a series of
baryonic root transitions on the nodes 3, 1, 3 in the order just specified. This yields the following sequence of cascading theories,
 \begin{equation}
 \label{cascadeU}
(2N,N,2N-1) \rightarrow  (2N,N,N+1) \rightarrow (1,N,N+1)\rightarrow (1,N,0)  \ .
 \end{equation}
This sequence could be thought as arising from the last three steps of an infinite cascade originated by an appropriate choice of fractional branes in the UV theory, in exactly the same way as the 
cascading theory of  \cite{Klebanov:2000hb} in the case of the conifold, or the cascading theory of 
\cite{Benini:2008ir,Cremonesi:2009hq} for the $ \mathbb{C}^2/\mathbb{Z}_2$ singularity. 
In our $\mathbb{C}^2/\mathbb{Z}_3 \times \mathbb{C}$ case, in analogy with $ \mathbb{C}^2/\mathbb{Z}_2$, every step in
 the duality cascade corresponds to the coupling of one of the gauge groups becoming very large; 
 this triggers the need to perform a baryonic root transition in order to continue the flow at lower energies using  a dual weakly coupled description.
 
Due to the strongly coupled nature of this flow, we are not justified to follow the 
computation order by order in the instanton number, and hence we should not be able to compare 
$k$-instantons of the theory at the $i$-th step of the cascade with $k$-instantons of the theory at the $(i-1)$-th step of the cascade. 
In particular, we should not be able to compare the gauge instanton computation of the theory just before the last step in the cascade 
and the stringy instanton computation of the theory after the last step in the cascade. However, the full prepotential should be continuous 
 along the cascade.  
In other terms, to follow the natural flow suggested by AdS/CFT correspondence,
 we should know the prepotential contributions at all  
instanton numbers for both theories.
On the contrary, in the weakly coupled flows we have chosen in the previous Sections, it was enough to compare the prepotentials 
order by order in the instanton number. 
It is nevertheless interesting to observe that the theory at the first step in the ``AdS/CFT'' cascade of Eq.~(\ref{cascadeU}), is exactly the 
UV theory we considered in Subsection \ref{Uflow}, namely $U(P+1) \times U(N)   \times U(P)$ (with $P=2N-1$).
The flow we discussed in that Subsection is controlled by the vacuum expectation value of $P$ diagonal components of the adjoint field of
the first node, and the VEV of the $P$ diagonal components of the adjoint field of the third node 
(see eqs (\ref{phi3vev},\ref{phi2vev})); 
these latter play the role of the masses for the hypermultiplets considered in Subsection \ref{Uflow}.

\begin{figure}[ht]
  \centering
  \includegraphics[scale=.29]{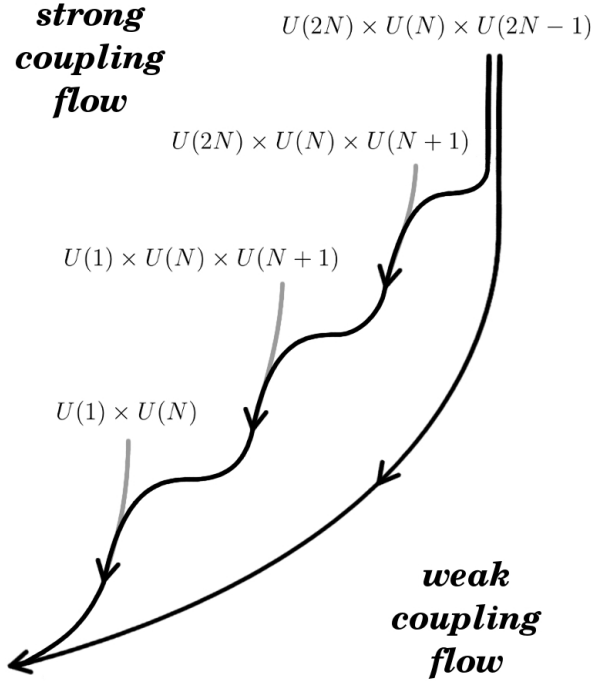}
  \caption{The last steps of an AdS/CFT inspired duality cascade and the interpolating weak coupling flow we use in the paper.}
  \label{flow}
\end{figure}

In \cite{Argyres:1996eh} it is shown that the particular weakly coupled flows we 
considered in the previous Sections lead exactly to the same IR effective theory reached after the last step 
in duality cascades such as the one in (\ref{cascadeU}). 
Hence we have the same UV completion for the IR theory as the one suggested by the AdS/CFT correspondence,
but, while the flow we followed is weakly coupled, the AdS/CFT flow passes through strongly coupled regions. 
The link between the ``AdS/CFT flow'' and the one we considered in Subsection \ref{Uflow} is explicitly shown in the cartoon in Figure \ref{flow}.  

The two different flows are controlled by different choices for the relations among the VEV of the adjoint hypermultiplets 
and the strong coupling scales of the gauge factors. As the prepotential is holomorphic, 
this choice should not induce any jump in the prepotential itself.
This observation based on holomorphicity closes the circle:
although we take an alternative journey that avoids the strong coupling regions of the flow,
we are actually computing and comparing non-perturbative effects at the various steps of the 
duality cascade associated to the natural flow provided by the AdS/CFT embedding of the theory.
Since we perform the comparison avoiding strongly coupled regions,
we can proceed order by order; in other terms, we are allowed to map
the gauge and the stringy instantons at every instanton number without needing to compute the full prepotential.

\subsection{Comments on $ \mathbb{C}^2/\mathbb{Z}_2$ and generalizations to $ \mathbb{C}^2/\mathbb{Z}_h$ }
\label{comments}

In this Subsection we focus on 
the stringy instanton contributions to unitary gauge theories in  both the particular example of 
$ \mathbb{C}^2/\mathbb{Z}_2$ and a simple generalization of the $ \mathbb{C}^2/\mathbb{Z}_h$ set up. 

The $\mathbb{C}^2/\mathbb{Z}_2$ quiver is special due to a doubling of the number of flavors.
Actually, the matter lines stretching between the two nodes are double lines.
In this case, although the computation proceeds in a similar way as the one we have already presented,
the final result is slightly different. 
Indeed, let us consider the example where the rank assignment is $(1,N)$: the number of colors is $N_c=1$, 
but the number of flavors is $N_f=2N$.
Eventually, the stringy $1$-instanton contribution is: 
\be
\label{Fk1C2Z2}
     \mathcal{F}^s_{k=1}= 
     M_s^{2-2 N}e^{2\pi i  \tau_{D}} \det \left(  \Phi_{2}-\Phi_1 {\bf 1}_{N}\right)^2 \ .
\ee
In analogy with the computations we have already described, it is possible to show that the stringy one-instanton
contribution is reproduced by the gauge one-instanton in the UV theory $(1+2P, N+P)$, i.e.~where here $N_c=1+2P$ and $N_f=2(N+P)$.\footnote{In this example, we note that the mass matrix that allows us to Higgs completely the $U(N_c)$ gauge group cannot be encoded in the adjoint field of the $U(N_f/2)$ group of the second node.}

A set up similar to the one that we have just analyzed for the $\mathbb{C}^2/\mathbb{Z}_2$ case 
provides a simple generalization for the generic $\mathbb{C}^2/\mathbb{Z}_h$ 
orbifold. Indeed, considering the rank assignment $(N_1, 1, N_3, 0,...,0)$,
the stringy $1$-instanton contribution at node 2 for this theory is
\be
\label{Fk1C2Zn}
     \mathcal{F}^s_{1}= 
     M_s^{2-N_1-N_3}e^{2\pi i  \tau_{D}} \det \left( \Phi_1- \Phi_{2}  {\bf 1}_{N_1}\right)
     \det \left(  \Phi_{3}-\Phi_2 {\bf 1}_{N_3}\right)\ .
\ee
Again, applying the same procedure we have previously used, it is possible to show
that the stringy instanton contribution \eqref{Fk1C2Zn} corresponds to the gauge 
$1$-instanton contribution of the UV theory $(N_1+P, 1+2P, N_3+P, 0,...,0)$.

\section{Conclusions}
We believe that we have closed the argument on the existence of truly
stringy instanton effects in four-dimensional theories engineered on
the world-volume 
of branes: the stringy instanton effects can be reconducted to field
theory instantonic effects. We showed this for ${\cal N}=2$ theories in
this paper. It was already shown for ${\cal N}=1$ theories that
stringy instanton effects could be ascribed to non-perturbative
effects, though generally strongly coupled. Actually, it is easy to
argue that also in  ${\cal N}=1$ set ups,  stringy
instanton effects can be related to gauge theory instantons in a weakly coupled way
by embedding into larger gauge theories, in complete analogy with the prescription presented in this paper. We discuss this issue in Appendix \ref{analogy}. 

In \cite{Billo:2009di} (see also \cite{Fucito:2009rs}), stringy instanton effects were computed in D$(-1)$/D7 brane set ups and  successfully mapped to the heterotic side up to instanton number 5. The complete sums of stringy instanton corrections were obtained in \cite{Petersson:2010qu} via 1-loop calculations of D-particles on the T-dual side and also of Horava-Witten BPS states on the M-theory side. Since we have argued that stringy instanton effects can be realized in terms of ordinary gauge instanton effects in UV completed gauge theories, this suggests direct relations between the various dual string theory perspectives and the Seiberg-Witten curve, as well as the Nekrasov formulae, for the corresponding gauge theory. As a future direction it would be interesting to explore these relations further.

Another direction is to consider the resummation of the instanton
series, in order to give a field theory interpretation of the
resulting expression for the prepotential. Such an expression could be
checked against instanton expansions also in theories that are related
by UV to IR flows passing through intermediate strongly coupled
regions, such as the cascade described in Subsection \ref{cascadeD}.

As a concluding remark, we hope that with this paper we have
underscored the deep interplay that occurs between instantonic effects
in string theory and in gauge theory, and the sort of complementary equivalence
that exists between the two pictures.

\section*{Acknowledgments}
 
We would like to thank 
Antonio Amariti,
Marco Bill\'o, Aldo Cotrone, Frank Ferrari, Gabriele Ferretti, Alberto Lerda, Diego Redigolo and Niclas Wyllard for important observations and suggestions.
The research of R.A., D.F., D.M. and C.P. is supported in part by IISN-Belgium (conventions
4.4511.06, 4.4505.86 and 4.4514.08), by the ``Communaut\'e
Fran\c{c}aise de Belgique" through the ARC program and by a ``Mandat d'Impulsion Scientifique" of the F.R.S.-FNRS. R.A. is a Research Associate and D.F. is a ``Charg\'e de recherches" of the Fonds de la Recherche Scientifique--F.R.S.-FNRS (Belgium). 
A.M. acknowledges funding by
the Durham International Junior Research Fellowship 
and by the
FWO PostDoctoral fellowship.
A. M. is also supported in part by
 the FWO-Vlaanderen through the project G.0114.10N, and in part by
the Belgian Federal Science Policy Office through the Interuniversity Attraction Pole
IAP VI/11.

\newpage

\appendix

\section{Masses for the symplectic quiver}
\label{sp_mass}

Using ${\cal N}=1$ notation, the superpotential of the ${\cal N}=2$ $U(N_1)\times U(N_2)\times {U}(N_3)$
gauge theory associated to the $\mathbb{C}^2/\mathbb{Z}_3 \times \mathbb{C}$
orbifold is given by
\begin{equation}\label{super_pot}
 \begin{split}
  \Phi^3_{(11)} (\Phi^1_{(12)} \Phi^2_{(21)} &- \Phi^2_{(13)} \Phi^1_{(31)})
 -\Phi^3_{(22)} (\Phi^2_{(21)} \Phi^1_{(12)} - \Phi^1_{(23)} \Phi^2_{(32)}) \\
 &+\Phi^3_{(33)} (\Phi^1_{(31)} \Phi^2_{(13)} - \Phi^2_{(32)} \Phi^1_{(23)})
 \end{split}
\end{equation}
Considering the orientifold specified in \cite{Ghorbani:2010ks}, we have
the following constraints
\begin{eqnarray}
 \Phi^3_{(11)} &= \epsilon\, (\Phi^3_{(11)})^T \epsilon \ , \
 \Phi^3_{(22)} &= - (\Phi^3_{(33)})^T \ , \\
 \Phi^1_{(12)} &= - \epsilon\, (\Phi^1_{(31)})^T \ , \
 \Phi^1_{(23)} &= (\Phi^1_{(23)})^T \ , \\
 \Phi^2_{(13)} &= \epsilon\, (\Phi^2_{(21)})^T \ , \
 \Phi^2_{(32)} &= (\Phi^2_{(32)})^T \ .
\end{eqnarray}
The implementation of the orientifold projection reduces the superpotential 
\eqref{super_pot} to
\begin{equation}
  2\ \Phi^3_{(11)} \Phi^1_{(12)} \Phi^2_{(21)} 
 -2\ \Phi^3_{(22)} (\Phi^2_{(21)} \Phi^1_{(12)} - \Phi^1_{(23)} \Phi^2_{(32)}) \ .
\end{equation}
We introduce here the following nomenclature
\begin{equation}
 \Phi = \Phi^3_{(11)} \ , \
 {\cal M} = \Phi^3_{(22)} \ , \
 X = \Phi^1_{(12)} \ , \
 Y = \Phi^2_{(21)} \ ,
\end{equation}
and we rewrite the part of the superpotential involving the node $1$ as follows,
\begin{equation}\label{super1}
  \Phi_I^{\ J} X_J^{\ \alpha} Y_\alpha^{\ I}  - {\cal M}_\alpha^{\ \beta} Y_\beta^{\ I} X_I^{\ \alpha} \ ,
\end{equation}
where the indexes $I,J$ run over $1,..., 2N_c$%
\footnote{We remind the reader that, in our notation, $Sp(N_c)$ has rank $N_c$ and
its fundamental representation has dimension $2 N_c$.}
and $\alpha,\beta$ run over $1,...,N_f$;
referring to the quiver depicted in Figure \ref{sympa}, we are here adopting
$N_1=N_c$ and $N_2=N_f$.
We split the gauge indices as follows
\begin{equation}
 I = (\underbrace{1,..,N_c}_i, \underbrace{N_c+1,..,2 N_c}_{\overline{i}} ) \ ;
\end{equation}
the matrix $\Phi$ is accordingly decomposed in four blocks
\begin{equation}
 \Phi_I^{\ J} = \left(\begin{array}{c|c}
                    \Phi_i^{\ j} & \Phi_i^{\ \overline{j}}\\ \hline 
                    \Phi_{\overline{i}}^{\ j} & \Phi_{\overline{i}}^{\ \overline{j}}
                   \end{array}\right) \ .                  
\end{equation}
We further take
\begin{equation}
 \Phi_I^{\ J} = \left(\begin{array}{cc}
                       1 & 0 \\ 0 & -1
                      \end{array}\right) \otimes \Phi \ ,
\end{equation}
with
\begin{equation}
 \Phi = \text{diag} (\phi_1, ..., \phi_{N_c}) \ ,
\end{equation}
so that
\begin{equation}
 \Phi_i^{\ j} = \phi_i\, \delta_i^{\ j} \ , \
 \Phi_i^{\ \overline{j}} =  \Phi_{\overline{i}}^{\ j} =0 \ , \
 \Phi_{\overline{i}}^{\ \overline{j}} = - \phi_{\,\overline{i}}\, \delta_{\overline{i}}^{\ \overline{j}} \ .
\end{equation}

We can rewrite \eqref{super1} as follows
\begin{equation}
  \left(\Phi_I^{\ J} \delta_{\alpha}^{\ \beta} - \delta_I^{\ J} {\cal M}_\alpha^{\ \beta} \right) X_J^{\ \alpha} Y_\beta^{\ I}
\end{equation}
where 
\begin{equation}
 \delta_I^{\ J} = \left(\begin{array}{c|c}
                    \delta_i^{\ j} & 0\\ \hline 
                    0 & \delta_{\overline{i}}^{\ \overline{j}}
                   \end{array}\right) \ .                 
\end{equation}
We then assign diagonal masses to the flavors,
\begin{equation}
 {\cal M}_\alpha^{\ \beta} = m_\alpha \delta_\alpha^{\ \beta}\ .
\end{equation}
Eventually, the superpotential \eqref{super1} is given by
\begin{equation}\label{super_fin}
 (\phi_i - m_\alpha) \left(X^{i\alpha} Y_{\alpha i} + X^{\overline{i}\alpha} Y_{\alpha \overline{i}}\right) \ .
\end{equation}

\section{Gauge theory $Sp(N_c)$ instanton computation}\label{Nekrasov}

In this Appendix we would like to give some details about the procedure to obtain the $k$-th order 
gauge instanton contribution to the prepotential.

The non-perturbative partition function
corresponding to the $k$-th instanton sector can be written as
\begin{equation}\label{integrand_fun}
 Z_k(\Phi; E_1,E_2) = \int d\vec{\chi}\ z_k(\vec{\chi};\Phi;E_1,E_2)\ ,
\end{equation}
where $\vec{\chi}$ represents the array of neutral moduli corresponding to the Cartan 
basis of the instanton group, $\Phi$ is the adjoint scalar field belonging to the vector
gauge supermultiplet and $E_1$, $E_2$ are the graviphoton background flux parameters.%
\footnote{\label{graviphoton} A technical though important ingredient which is essential to employ
localization techniques is the so-called $\Omega$-deformation; it
essentially amounts to introducing a non-vanishing background flux
for the graviphoton \cite{Billo:2006jm}.
The presence of a non-null graviphoton flux regularizes the integral
over the instanton moduli parameterizing the instanton center position
in superspace. The graviphoton background corresponds to a closed
RR flux and introduces a non-zero curvature for the superspace.
The final results obtained through localization, namely the instanton
partition function and the instanton contribution to the prepotential,
are well defined also in the limit of zero graviphoton flux, i.e.
in flat superspace. On the computational level, the graviphoton background
flux 
is parametrized by two quantities, namely $E_1$ and $E_2$, corresponding
to the two Cartan directions of the $4$-dimensional Lorentz group.
The instanton contributions in flat superspace are obtained considering the limit 
$E_1,E_2 \rightarrow 0$.} The function $z_k(\vec{\chi};\Phi;E_1,E_2)$
encodes the result of the integration over all the instanton moduli but $\vec{\chi}$
of the exponentiated instanton action multiplied by the Vandermonde factor
(introduced when considering the matrix ${\chi}$ along its Cartan, see 
for instance \cite{Ghorbani:2010ks}).

In the presence of matter, the integrand function \eqref{integrand_fun}
admits the following factorization
\begin{equation}\label{factorization}
 z_k(\vec{\chi};\Phi;E_1,E_2) = z_k^{(\text{pure gauge})}(\vec{\chi};\Phi;E_1,E_2)\ \cdot \ z_k^{(\text{matter})}(\vec{\chi};\Phi;E_1,E_2)\ .
\end{equation}
In the case of $Sp(N)$ gauge group, the ordinary $k$-instanton symmetry group 
is $SO(k)$ whose rank is $[k/2]$; this means that, at level $k$, we have $[k/2]$
Cartan directions parameterized by $\vec{\chi}$. Referring to \cite{Shadchin:2004yx,Shadchin:2005mx},
the pure gauge factor in \eqref{factorization} is given by
\begin{equation}
 \begin{split}
 &z_k^{(\text{pure gauge})}(\vec{\chi};\Phi;E_1,E_2) = \frac{(-1)^{[k/2]+\delta}}{2^{[k/2]+\delta}[k/2]!}\left(\frac{E_1+E_2}{E_1 E_2}\right)^{[k/2]} \frac{\Delta(0) \Delta(E_1+E_2)}{\Delta(E_1)\Delta(E_2)}\\
 &\ \ \ \  \prod_{i=1}^{[k/2]} \frac{1}{P[\chi_i-
       (E_1+E_2)/2]\ P[\chi_i+ (E_1+E_2)/2]\ (4\chi_i^2 - E_1^2)
     (4\chi_i^2 - E_2^2)}\\
&\ \ \ \ \left(\frac{1}{E_1E_2P[(E_1+E_2)/2]}\prod_{i=1}^{[k/2]}\frac{\chi_i^2(\chi_i^2-(E_1+E_2)^2)}{(\chi_i^2-E_1^2)(\chi_i^2-E_2^2)}\right)^\delta
 \end{split}
\end{equation}
where we have defined
\begin{eqnarray}
 \Delta(x) &=& \prod_{1\leq i< j \leq [k/2]} \left[(\chi_i + \chi_j)^2 - x^2\right] \left[(\chi_i - \chi_j)^2 - x^2\right] \ , \\ \label{P}
 P[x] &=& \prod_{l=1}^{N} (x^2 - \phi_l^2) \ ,
\end{eqnarray}
and $\delta=0$ when $k$ is even, while $\delta=1$ when $k$ is odd.

In the presence of $N_f$ fundamental flavors, the matter factor in $z_k$ is 
given by (see again \cite{Shadchin:2004yx,Shadchin:2005mx})
\begin{equation}
 z_k^{(\text{matter})}(\vec{\chi};\Phi;E_1,E_2) = \prod_{j=1}^{N_f} \left(m_j-\frac{1}{2}(E_1 + E_2)\right)^\delta\prod_{i=1}^{[k/2]} \left[\left(m_j - \frac{1}{2}(E_1 + E_2)\right)^2 - \chi_i^2\right] \ .
\end{equation}

Eventually, in order to obtain the partition functions $Z_k$ in \eqref{integrand_fun},
we have to integrate over the instanton moduli $\vec{\chi}$; following Nekrasov's 
prescription, the integration variables $\chi_i$ must be complexified one at a time and
the integral itself has to be performed via the residues method.
Nekrasov's prescription requires that the singularities encountered along the real $\chi_i$
axes be slightly moved away from it assigning appropriate, small complex displacements; 
moreover, the integration contours are closed at infinity even though the integrand
function does not go to zero at large distances from the origin. Even though not completely
motivated \emph{a priori}, Nekrasov's prescription has been widely tested and proved to
lead to correct results whenever alternative methods are available.

We skip the detail of the $\vec{\chi}$ integration and jump to the final results.
The relation between the non-perturbative prepotential and the
instanton partition function is (see for instance \cite{Billo':2010bd})
\begin{equation}\label{prep_part}
 F^g= (E_1 E_2)\ \log Z\ .
\end{equation}
Both sides of \eqref{prep_part} can be expanded according to the topological charge $k$ obtaining
\begin{eqnarray}
 F_1^g &=& (E_1 E_2) Z_1\ ,\\
 F_2^g &=& (E_1 E_2) Z_2 - \frac{(F_1^g)^2}{2 (E_1 E_2)}\ ,\\
 F_3^g &=& (E_1 E_2) Z_3 - \frac{F_1^g F_2^g}{(E_1 E_2)} - \frac{(F_1^g)^3}{6 (E_1 E_2)^2}\ .
\end{eqnarray}
Finally, the explicit results up to instanton number three are
reported in Eqs.~\eqref{effe1}--\eqref{T}, where they are already
expressed in the flat spacetime limit of vanishing graviphoton
background, $E_1,E_2 \rightarrow 0$.

\section{Instanton partition function regularizations}
\label{regu}

In this Appendix we evaluate the integral \eqref{I} by using three different regularizations and show that they give the same result.

\subsection{VEV regularization}
\label{vevreg}

\begin{eqnarray}
\label{ad}
 \mathcal{I}_{v}&=&\int  d^2\omega_{\dot\alpha} d^2 \ov\omega^{\dot\alpha}d^3D^c~  \overline{\omega}^{\dot\alpha}  \omega_{\dot\alpha} \,
   e^{ -i  D^c\, \overline{\omega}^{\dot{\alpha}} (\tau^c)_{\dot{\alpha}}^{\dot{\beta}} \omega_{\dot{\beta}}
   -\frac{1}{2} |v|^2 \overline{\omega}^{\dot{\alpha}}  \omega_{\dot{\alpha}}} \nn \\
   &=& \int  d^2\omega_{\dot\alpha} d^2 \ov\omega^{\dot\alpha}d^3D^c~ \left(-\frac{\partial}{\partial M_1}-\frac{\partial}{\partial M_4}\right)\, \text{exp}\left\{ - (\overline{\omega}^{\dot{1}}, \overline{\omega}^{\dot{2}}) 
        \left(\begin{array}{cc}
         M_1 & M_2 \\ M_3 & M_4
        \end{array}\right)
        \left(\begin{array}{c}
               \omega_{\dot{1}}\\
               \omega_{\dot{2}}\\
              \end{array}\right)  \right\} \nn \\
&=&4\pi^2   \int d^3D^c  \left(-\frac{\partial}{\partial M_1}-\frac{\partial}{\partial M_4}\right)
      \frac{1}{M_1 M_4 - M_2 M_3}\nn \\
 &=&4\pi^2  \int d^3D^c
      \frac{|v|^2}{[\frac{1}{4}|v|^4 + |D^c|^2]^2} = 16\pi^3  \int_0^\infty d D\ D^2  
 \frac{|v|^2}{[\frac{1}{4}|v|^4 + D^2]^2}           \nn \\
 &=&32\pi^3  \int_0^\infty d \tilde{D}\ \tilde{D}^2 \frac{1}{(1 +
   \tilde{D}^2)^2} = 8\pi^4\ ,
\end{eqnarray}
where $ M_1 =  i D^3 + |v|^2/2$, $ M_2 = i (D^1 +i D^2)$, 
 $ M_3 = i (D^1 - i D^2)$, $ M_4 = - i D^3 + |v|^2/2$, we have rewritten the $D^c$ variables in terms of spherical coordinates and we have performed the rescaling $\tilde{D} = D/(4|v|^2)$.

\subsection{Non-commutative regularization}
\label{nonreg}

\begin{eqnarray}
\mathcal{I}_{\xi}&=&  \int  d^2\omega_{\dot\alpha} d^2 \ov\omega^{\dot\alpha}d^3D^c~  \ov\omega^{\dot\alpha}\omega_{\dot\alpha}\,e^{ -i D^c \ov\omega^{\dot \alpha}
(\tau^c)^{\dot\beta}_{\dot\alpha} \omega_{\dot\beta} +i D^c\xi\delta_{c3} }\nn \\
&=&(2\pi)^3 \int  d^2\omega_{\dot\alpha} d^2 \ov\omega^{\dot\alpha}~  \ov\omega^{\dot\alpha}\omega_{\dot\alpha}\,\prod_c \delta \left( \ov\omega^{\dot \alpha}
(\tau^c)^{\dot\beta}_{\dot\alpha} \omega_{\dot\beta} - D^c\xi\delta_{c3} \right) \nn \\
&=&(2\pi)^3 \int  d^4 y  \left( \vec{y}\cdot \vec{y}  \right)
\delta \left( y_1 y_3+ y_2 y_4\right)
\delta \left( y_1 y_4- y_2 y_3 \right)
\delta \left( y_1^2+y_2^2-y_3^2-y_4^2- \xi \right)\nn \\
&=&(2\pi)^3 \int d^3 y  \left( y_1^2+y_2^2+y_3^2+ \frac{y_1^2 y_3^2}{y_2^2} \right) \frac{\delta (y_3)}{y_1^2+y_2^2}\,
\delta \left( y_1^2+y_2^2+y_3^2+\frac{y_1^2 y_3^2}{y_2^2} - \xi \right)\nn \\
&=&(2\pi)^3 \int r \,dr\, d\theta\, \frac{\delta \left( r - \sqrt{\xi}
  \right)}{2r} = 8\pi^4\ ,
\end{eqnarray}
where we have rewritten the $\omega_{\dot\alpha}$ and
$\ov\omega^{\dot\alpha}$ variables first in terms of real coordinates,
$\omega_{\dot{1}} = y_1+i y_2$ and $\omega_{\dot{2}}=y_3+iy_4$ (note
that there is a factor of 4 from the Jacobian) and then in terms of polar ones.

\subsection{$\alpha^\prime$ regularization}
\label{areg}

\begin{eqnarray}
\mathcal{I}_{\alpha^\prime}&=&  \int  d^2\omega_{\dot\alpha} d^2 \ov\omega^{\dot\alpha}d^3D^c~  \ov\omega^{\dot\alpha}\omega_{\dot\alpha}\,e^{ -i D^c \ov\omega^{\dot \alpha}
(\tau^c)^{\dot\beta}_{\dot\alpha} \omega_{\dot\beta} -\frac{1}{2g_{0}^{2}}  (D^c)^2} \nn \\
&=&4 \int  d^3D^c d^4 y
~\left( \vec{y}\cdot \vec{y}  \right)~e^{-2i (y_1 y_3+y_2 y_4)D^1- \frac{1}{2g_{0}^{2}} (D^1)^2 }
~
 \nn \\
&& \qquad \quad \times ~ e^{-2i (y_1 y_4-y_2y_3)D^2- \frac{1}{2g_{0}^{2}} (D^2)^2} ~~ e^{-i (y_{1}^{2}+y_{2}^{2}-y_{3}^{2}-y_{4}^{2})D^3- \frac{1}{2g_{0}^{2}} (D^3)^2 }\nn \\
&=& 4\big(2\pi g_{0}^{2}\big)^{\frac{3}{2}} \int   d^4 y ~\left( \vec{y}\cdot \vec{y}  \right)  ~e^{-\frac{g_{0}^{2}}{2}(\vec{y}\cdot \vec{y})^2}\nn \\
&=& 
4\big(2\pi g_{0}^{2}\big)^{\frac{3}{2}}(\mathrm{vol}S^3) \int d\rho
~\rho^5 ~e^{-\frac{g_{0}^{2}}{2}\rho^4} =8\pi^{4}\ .
\end{eqnarray}

\section{Analogy with the $\mathcal{N}=1$ case}
\label{analogy}

The correspondence between stringy instantons and gauge instantons in $\mathcal{N}=1$
gauge theories has been already explored in \cite{GarciaEtxebarria:2007zv,Aharony:2007pr,Krefl:2008gs,Amariti:2008xu}.
We here review in a simple quiver-like example some aspects which
 clarify the analogy with the $\mathcal{N}=2$ case that we have studied in this paper.
Indeed, also in the $\mathcal{N}=1$ case, the stringy instanton in the IR theory can be understood 
as an ordinary gauge instanton in a UV completion of the theory. 
The flow between the UV and the IR theory involves Higgsing of the gauge group and at the same
time integrating out some flavors.

\begin{figure}[ht]
  \centering
  \includegraphics[scale=.26]{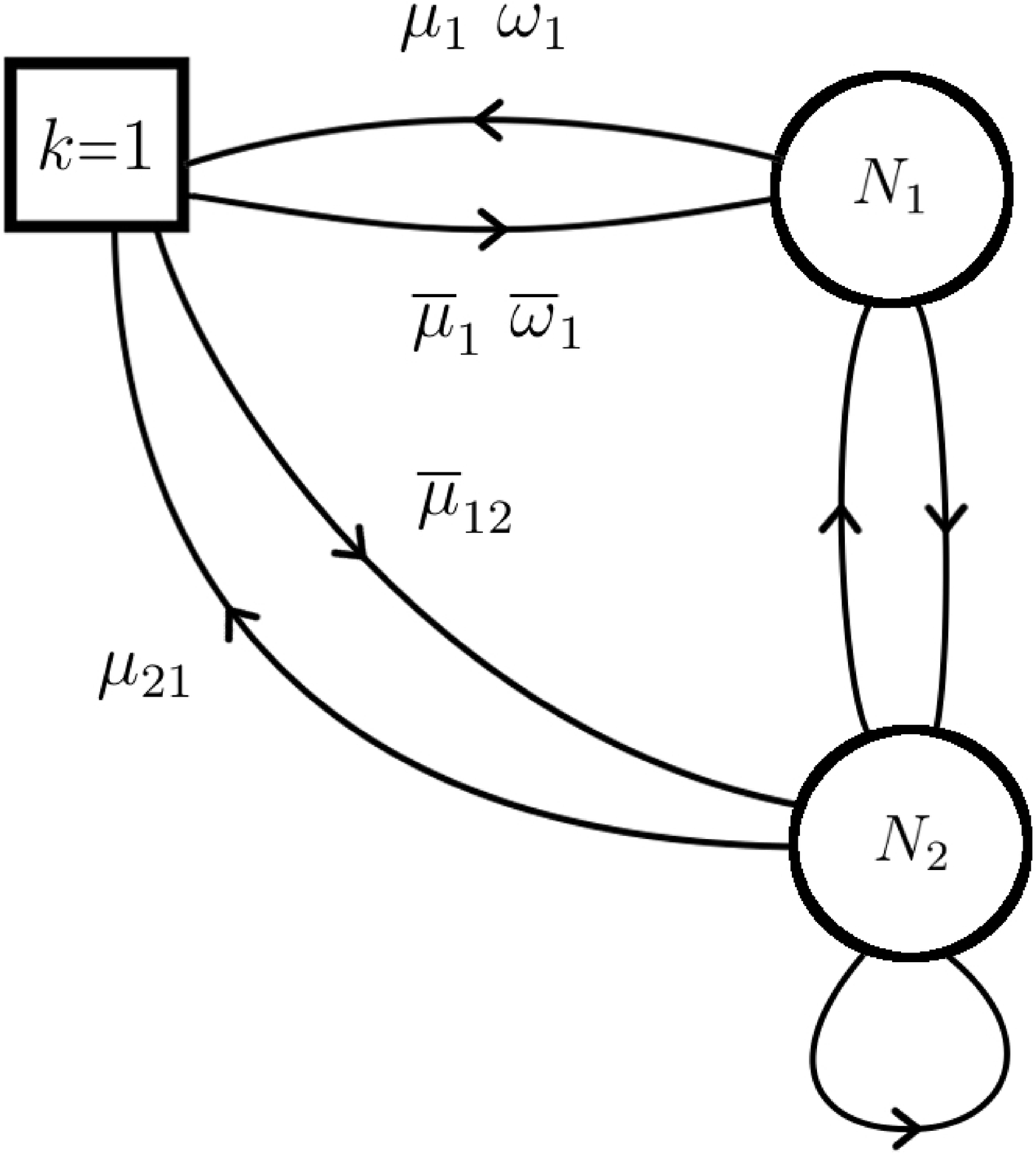}
  \caption{The $\mathcal{N}=1$ $U(N_1) \times U(N_2)$ quiver gauge theory with one D-brane instanton on node $1$.}
  \label{quivermariot}
\end{figure}
Consider the quiver in the Figure \ref{quivermariot}, with $N_1=1$, $N_2=N$ and with superpotential
\be
\label{superIR}
W=\hat X q_{21} q_{12}\ .
\ee
A D$(-1)$ stringy instanton on node $1$ gives the following contribution to the superpotential, 
up to a numerical factor,
\be
\label{instaN1}
W^s_{inst}= M_s^{3-N} e^{2 \pi i \tau_D} \det \hat X\ .
\ee

The same contribution can be obtained through a proper gauge instanton computation
as follows.
We consider the UV completion of the previous theory as the $U(N_1=1+P)\times U(N_2=N+P)$
quiver gauge theory with superpotential
\be
W=X_{22}Q_{21}Q_{12}+m^2 Z\ ,
\ee
where we have split the field $X$ as
\be
X_{22}=\left(
\begin{array}{cc}
Z & Y \\
\tilde Y & \hat X
\end{array}
\right)\ .
\ee
Here $Z$ is a $P \times P$ square matrix and $\hat X$ is an $N \times N$ square matrix.
On the vacuum of this theory, the first $P$ entries of the fields $Q_{21}$ and $Q_{12}$ get
a VEV equal to $m$. 
The gauge group $U(1+P)$ is Higgsed down to $U(1)$ and the flavors are reduced from $N+P$ to $N$.
The low energy scale is defined as
\be
\label{lowscale}
\Lambda_{IR}^{3-N}=\frac{\Lambda_{UV}^{3+ 2 P- N}}{m^{2 P}}\ .
\ee
The IR theory is the one discussed above with superpotential \eqref{superIR},
where $q_{12}$ and $q_{21}$ are the remaining parts of $Q_{21}$ and $Q_{12}$
that do not get a VEV.

Now we compute the gauge instanton contribution to the superpotential
in the UV theory. 
The instanton action for one instanton
placed on node $1$ is
\bea
&&S_{inst}=S_1+S_2+S_W \\
&& S_1=i ( \bar \mu_1 \omega_{\dot \alpha 1}+\bar \omega_{\dot \alpha 1} \mu_1) \lambda^{\dot \alpha}
-i D^c(\bar \omega_{\dot \alpha 1} \tau^c \omega_{\dot \alpha 1}) \nonumber \\
&& S_2 =  \bar \omega_{1} Q_{21}^{\dagger}Q_{21} \omega_1+\bar \omega_1 Q_{12} Q_{12}^{\dagger} \omega_1
+i \bar \mu_{12} Q_{12}^{\dagger} \mu_1- i \bar \mu_1 Q_{21}^{\dagger} \mu_{21} \nonumber \\
&& S_W=-i \bar \mu_{12} X_{22} \mu_{21}\ . \nonumber
\eea
The integral over the instanton moduli space gives
\be
\mathcal{Z}=\Lambda_{UV}^{3+2P-N} \int d a_\mu d M dD d\omega_1 d\bar \omega_1 d\lambda  d\mu_1^{N_1} d\bar \mu_1^{N_1} 
d \bar \mu_{12}^{N_2} d \mu_{21}^{N_2} e^{-S_{inst}}\ .
\ee
The measure $d a_\mu d M$ is interpreted as the $\mathcal{N}=1$ superspace measure, hence giving the following
contribution to the superpotential
\be
W_{inst}^g=\Lambda_{UV}^{3+2P-N} \int dD d\omega_1 d\bar \omega_1 d\lambda  d\mu_1^{N_1} d\bar \mu_1^{N_1} 
d \bar \mu_{12}^{N_2} d \mu_{21}^{N_2} e^{-S_{inst}}\ .
\ee
We focus now on the fermionic integrations. The integral over $\lambda$ can be done using the $S_1$ part of the action and
gives the usual ADHM constraints
\be
W_{inst}^g \propto \Lambda_{UV}^{3+2P-N} \int  d\mu_1^{N_1} d\bar \mu_1^{N_1} 
d \bar \mu_{12}^{N_2} d \mu_{21}^{N_2} \left(\bar \mu_1 \omega_1+\bar
\omega_1 \mu_1 \right)^2 e^{-S_{2}-S_{W}}\ ;
\ee
this soaks up only one of each of the fermionic zero modes $\mu_1$ and $\bar \mu_1$.
Now, we saturate the remaining $(N_1-1)$ $\mu_1$ and $(N_1-1)$ $\bar \mu_1$ 
fermionic zero modes by using the action $S_2$
\bea
W_{inst}^g&\propto&\Lambda_{UV}^{3+2P-N} \int  d\mu_1^{N_1} d\bar \mu_1^{N_1} 
d \bar \mu_{12}^{N_2} d \mu_{21}^{N_2} 
\left(\bar \mu_1 \omega_1+\bar \omega_1 \mu_1 \right)^2 \cdot \nonumber \\
&& \qquad \qquad \qquad \quad
\cdot (\bar \mu_{12} Q_{12}^{\dagger}\mu_1 )^{N_1-1} (\bar \mu_1 Q_{21}^{\dagger} \mu_{21})^{N_1-1}
e^{-S_{W}}\ .
\eea
This also saturates $N_1-1$ of $\bar \mu_{12}$ and $N_1-1$ of $\mu_{21}$ fermionic zero modes.
Finally, we saturate the remaining $N_2-N_1+1$ of $\bar \mu_{12}$ and $N_2-N_1+1$ of $\mu_{21}$ fermionic zero modes
by using the action $S_W$, obtaining
\bea
W_{inst}^g&\propto&
\Lambda_{UV}^{3+2P-N}  \int  d\mu_1^{N_1} d\bar \mu_1^{N_1} 
d \bar \mu_{12}^{N_2} d \mu_{21}^{N_2} \left(\bar \mu_1 \omega_1+\bar \omega_1 \mu_1 \right)^2 \cdot \\
&& 
 \qquad \qquad \qquad 
 \cdot
(\bar \mu_{12} Q_{12}^{\dagger}\mu_1 )^{N_1-1} (\bar \mu_1 Q_{21}^{\dagger} \mu_{21})^{N_1-1}
(\bar \mu_{12} X_{22} \mu_{21})^{N_2-N_1+1}\ . \nonumber
\eea
This procedure saturates all the fermionic zero modes and can give a non vanishing result.
Note that we can split the integration in two independent pieces
\bea
W_{inst}^g&\propto&\int  d\mu_1^{N_1} d\bar \mu_1^{N_1} 
d \bar \mu_{12}^{N_1-1} d \mu_{21}^{N_1-1}
\left(\bar \mu_1 \omega_1+\bar \omega_1 \mu_1 \right)^2 
(\bar \mu_{12} Q_{12}^{\dagger}\mu_1 )^{N_1-1} (\bar \mu_1 Q_{21}^{\dagger} \mu_{21})^{N_1-1}  \nonumber \\
&& \qquad \cdot ~
\Lambda_{UV}^{3+2P-N}
\label{final1}
\int d \bar \mu_{12}^{N_2-N_1+1} d \mu_{21}^{N_2-N_1+1}
(\bar \mu_{12} X_{22} \mu_{21})^{N_2-N_1+1}\ .
\eea
The first one, re-including also the bosonic integration, is the same as the one discussed in 
\cite{Argurio:2007vqa} and gives the same result. The second part gives a sub-determinant of
the field $X_{22}$. 
Note that in the vacuum only the first $N_1-1=P$ components of $Q_{12}$ and $Q_{21}$ get
a VEV. 
So in order to get a non-vanishing contribution we consider to have selected precisely
those components in the 
first line of (\ref{final1}); as a consequence in the second line only the $\hat X$ part of the field $X_{22}$ 
is involved (remember that $N_2-N_1+1=N$).

We conclude that the instanton contribution to the superpotential is
\be
W_{inst}^g= \frac{\Lambda_{UV}^{3+2P-N}}{\langle Q_{12} \rangle^{N_1-1}  \langle Q_{12} \rangle^{N_1-1}} \det \hat X=
 \frac{\Lambda_{UV}^{3+2P-N}}{m^{2 N_1-2}} \det \hat X\ ,
\ee
where here in the second passage we have inserted the VEVs
for the fields $Q_{12}$ and $Q_{21}$.
Remembering that $N_1=1+P$, $N_2=P+N$,
and matching the string scale to the low energy scale \eqref{lowscale}
as
\be
M_s^{3-N} e^{2 \pi i \tau_D} =\Lambda_{IR}^{3-N}\ ,
\ee
we recover exactly the contribution \eqref{instaN1}.

This procedure 
provides a gauge theory interpretation of the stringy instantons
in $\mathcal{N}=1$ theories which is analogous to the approach we have
adopted for the $\mathcal{N}=2$ case.

Note that the instanton computation leading to the contribution \eqref{instaN1} for an s-confining 
gauge theory has been discussed in \cite{Seiberg:1994pq} by 
interpreting the low energy description of the $N_f=N_c+1$ theory as
a completely Higgsed $SU(2)$ magnetic theory with $N_f+1$ flavors.
Here we have extended the
computation to an arbitrary number $P$ of integrated-in flavors in order to make a connection
with the $\mathcal{N}=2$ case.

\end{document}